\documentclass[referee]{aa}

\usepackage{hyperref}
\usepackage{amsmath}
\usepackage{graphicx}
\usepackage{appendix}
\usepackage{txfonts}
\begin{document}

   \title{Formation of N--bearing complex organic molecules in molecular clouds: Ketenimine, acetonitrile, acetaldimine, and vinylamine via the UV photolysis of C$_2$H$_2$ ice}

    \titlerunning{Formation of N--bearing complex organic molecules}
   
   \author{
   K.-J. Chuang\inst{1,2}
   \and C. Jäger\inst{1}
   \and J. C. Santos\inst{2}
   \and Th. Henning\inst{3}
    }
\authorrunning{Chuang et al.}
   \institute{Laboratory Astrophysics Group of the Max Planck Institute for Astronomy at the Friedrich Schiller University Jena, Institute of Solid State Physics, Helmholtzweg 3, D-07743 Jena, Germany\\
   \and
   Laboratory for Astrophysics, Leiden Observatory, Leiden University, P.O. Box 9513, NL-2300 RA Leiden, the Netherlands.
   \email{chuang@strw.leidenuniv.nl}
   \and
    Max Planck Institute for Astronomy, Königstuhl 17, 69117 Heidelberg, Germany}        

  \date{}


  \abstract
   {The solid-state C$_2$H$_2$ chemistry in interstellar H$_2$O-rich ice has been proposed to explain astronomically observed complex organic molecules (COMs), including ketene (CH$_2$CO), acetaldehyde (CH$_3$CHO), and ethanol (CH$_3$CH$_2$OH), toward early star-forming regions. This formation mechanism is supported by recent laboratory studies and theoretical calculations for the reactions of C$_2$H$_2$+OH/H. However, the analog reaction of C$_2$H$_2$+NH$_2$ forming N-bearing species has been suggested to have a relatively low rate constant that is orders of magnitude lower than the value of C$_2$H$_2$+OH.}
   {This work extends our previous laboratory studies on O-bearing COM formation to investigate the interactions between C$_2$H$_2$ and NH$_3$ ice triggered by cosmic ray-induced secondary UV photons under molecular cloud conditions.}
   {Experiments were performed in an ultra-high vacuum chamber to investigate the UV photolysis of the C$_2$H$_2$:NH$_3$ ice mixture at 10 K. The ongoing chemistry was monitored in situ by Fourier-transform infrared spectroscopy as a function of photon fluence. The IR spectral identification of the newly formed N-bearing products was further secured by a quadrupole mass spectrometer during the temperature-programmed desorption experiment.}
   {The studied ice chemistry of C$_2$H$_2$ with NH$_2$ radicals and H atoms resulting from the UV photodissociation of NH$_3$ leads to the formation of several N-bearing COMs, including vinylamine (CH$_2$CHNH$_2$), acetaldimine (CH$_3$CHNH), acetonitrile (CH$_3$CN), ketenimine (CH$_2$CNH), and tentatively ethylamine (CH$_3$CH$_2$NH$_2$). The experimental results show an immediate and abundant CH$_2$CHNH$_2$ yield as the first-generation product, which is further converted into other chemical derivatives. The effective destruction and formation cross-section values of parent species and COMs were derived, and we discuss the chemical links among these molecules and their astronomical relevance.}
   {}


   \maketitle
%

\section{Introduction}

Astronomical observations have revealed a rich molecular inventory in star-forming regions, including simple volatiles (such as H$_2$O, CO, CH$_4$, NH$_3$) and complex organics consisting of molecules with six or more atoms \citep{Herbst2009, Boogert2015, McGuire2018}. These interstellar molecules have been proposed to be further incorporated into part of planetesimals or cometesimals, ultimately enriching the molecular complexity in planetary systems. The (partial) linear correlation between molecules detected in the protostar IRAS 16923-2422 and the cometary material discovered by the Rosetta Mission toward the comet 67P/C-G supports the concept of the inherited molecular evolution \citep{Drozdovskaya2019}. For example, oxygen--bearing complex organic molecules (COMs) have been unambiguously found toward high- and low-mass protostars, and even in several prestellar stages, such as B1-b, L1689B, and B5 \citep{Bisschop2008, Oberg2010, Bacmann2012, Cernicharo2012, Taquet2017}. Among them, the simultaneous detection of O--bearing COMs characterized by a degree of hydrogenation, such as C$_2$H$_n$O$_2$ (e.g., glycolaldehyde and ethylene glycol) and C$_2$H$_n$O (e.g., ketene, acetaldehyde, and ethanol) hint at their solid-state chemical history through hydrogen addition reactions on grain surfaces \citep{Jorgensen2012, Jorgensen2016, vanGelder2020}. Besides O--bearing COMs, nitrogen-containing organics (i.e., H$_n$-NC-R) are also considered essential ingredients in the formation of prebiotic compounds such as amino acids and nucleobases. Several N--bearing COMs expressed by the formula C$_2$H$_n$N (n=3 and 5), including vinylamine (CH$_2$CHNH$_2$), acetaldimine (CH$_3$CHNH), ketenimine (CH$_2$CNH), and acetonitrile (CH$_3$CN) have been identified in giant molecular clouds toward Sgr B2 \citep{Lovas2006, Loomis2013, Thiel2017, Zeng2021}. However, the fully saturated species ethylamine (CH$_3$CH$_2$NH$_2$) remains a tentative detection \citep{Zeng2021}.

\begin{table*}[t]
    \centering
    \caption{Summary of band strength values used in this work.}
    \begin{tabular}{lccccc}
    \hline
    \hline
    Species & Chemical formula & IR peak position & Band strength value & References & Note \\
    & & (cm$^{-1}$) & (cm molecule$^{-1}$) & & \\
    \hline
    Acetylene & C$_2$H$_2$ & 810 & 2.40$\times$10$^{-17}$ & \cite{Hudson2014} & exp. \\
    Ammonia & NH$_3$ & 1067 & 1.95$\times$10$^{-17}$ & \cite{Hudson2022} & exp. \\
    Hydrogen cyanide & HCN & 2107 & 1.03$\times$10$^{-17}$ & \cite{Gerakines2022} & exp. \\
    Cyanid anion & CN$^{-}$ & 2081 & 3.70$\times$10$^{-17}$ & \cite{Georgieva2006} & cal. \\
    Acetonitrile & CH$_3$CN & 2253 & 2.20$\times$10$^{-18}$ & \cite{Hudson2004} & exp. \\
    Ketenimine & CH$_2$CNH & 2054 & 7.20$\times$10$^{-17}$ & \cite{Hudson2004} & est. \\
    Vinylamine & CH$_2$CHNH$_2$ & 1302 & 1.20$\times$10$^{-18}$ & \cite{Lammertsma1994} & cal. \\
    Acetaldeimine & CH$_3$CHNH & 1403 & 2.80$\times$10$^{-18}$ & \cite{Lammertsma1994} & cal. \\
    \hline
    \end{tabular}
    \label{table01}
\end{table*}

Several gas-phase and solid-phase formation pathways have been proposed to explain the above observations, particularly for O--bearing COMs, either through “bottom-up” (starting with CO or C atoms) or “top-down” (starting with fragments of carbonaceous species) routes \citep{Tielens2013}. In the solid-state scenario, atomic species or simple molecules accrete onto cold dust grains in molecular clouds, facilitating the surface association to form larger molecules. For example, CO hydrogenation not only results in H$_2$CO and CH$_3$OH but also offers key radicals, namely HCO, CH$_2$OH, and CH$_3$O for O--bearing COM formation, such as CH$_2$OHCHO and CH$_2$OHCH$_2$OH, through the radical-radical reaction mechanism \citep{Butscher2015, Fedoseev2017, Chuang2016, Chuang2017, Simons2020, He2022, Tsuge2023}. The intermediate species also react with other radicals or participant gas-phase reactions through non-thermal desorption, leading to heterogeneous COMs \citep{Vazart2020}. On the other hand, highly unsaturated hydrocarbon alkynes, such as C$_2$H$_2$, which could originate from PAH photodissociation, gas-phase production, or even sublimation of carbonaceous materials, can also associate with OH/H radicals to form O--bearing COMs. The solid-state reactions based on the --C$\equiv$C-- molecular backbone have been investigated in theoretical and experimental works \citep{Tielens1992, Hudson1997, Hiraoka2000, Wu2002, Kobayashi2017}. Recently, \cite{Molpeceres2022} systematically studied the interactions of C$_2$H$_n$ (n=2, 4, and 6) with several radicals, including H, OH, NH$_2$, and CH$_3$, and reported the prevalence of OH--bearing products (e.g., C$_2$H$_2$OH and C$_2$H$_4$OH) on grain surfaces. In addition, the association of C$_2$H with H$_2$O has been proposed to form C$_2$H$_2$OH through a barrierless reaction \citep{Perrero2022}.

Our previous laboratory studies reported on the solid-state COM formation based on C$_2$H$_2$ ice under molecular cloud conditions through non-energetic (i.e., OH/H addition reactions) and energetic processing (i.e., proton radiolysis of H$_2$O) \citep{Chuang2020, Chuang2021}. The direct attachment of OH radicals to C$_2$H$_2$ has been suggested to efficiently form the isomers {vinyl alcohol} (CH$_2$CHOH) and acetaldehyde (CH$_3$CHO). These newly formed O--bearing COMs could further react with H atoms on dust grains, leading to {ketene} (H$_2$CCO) and ethanol (CH$_3$CH$_2$OH) through hydrogen abstraction and addition reactions, respectively. The experimental findings align with the theoretical work of Molpeceres and Rivilla (2022), which shows a relatively low activation barrier of 2.39 kcal mol$^{-1}$ for the reaction C$_2$H$_2$+OH$\rightarrow$C$_2$H$_2$OH. Nevertheless, this theoretical work reports that the contribution of the analog reactions of C$_2$H$_2$ with NH$_2$ and CH$_3$ radicals are minor (see Figure 1 in \citealt{Molpeceres2022}). The calculated activation barrier of C$_2$H$_2$+NH$_2$$\rightarrow$C$_2$H$_2$NH$_2$ is only somewhat high (i.e., 6.93 kcal mol$^{-1}$), but the reaction rate constant (i.e., $\sim$10$^{-6}$ s$^{-1}$) in the solid state is several orders of magnitude lower than those for the reactions C$_2$H$_2$+OH$\rightarrow$C$_2$H$_2$OH and C$_2$H$_2$+H$\rightarrow$C$_2$H$_3$. It is even less possible for the reaction of C$_2$H$_2$+CH$_3$$\rightarrow$C$_2$H$_2$CH$_3$ (i.e., $\sim$10$^{-13}$ s$^{-1}$). This leads toward the question of whether the so-called top-down mechanism based on C$_2$H$_2$ ice can lead to N--bearing COMs on dust grains like its O--bearing counterpart. In laboratory studies, a 5-keV electron beam has been used to simulate the cosmic ray-induced ice chemistry of C$_2$H$_2$ and NH$_3$ ice, showing the formation of C$_2$H$_3$N isomers, such as ethynamine (HCCNH$_2$) and 2H-azirine(c-H$_2$CCHN), and CH$_2$CHNH$_2$ at $\sim$5 and 10 K, respectively \citep{Turner2021,Zhang2023}. \cite{Volosatova2022} reported the high-resolution IR spectroscopy of C$_2$H$_2$:NH$_3$ in noble gas (Ar, Kr, or Xe) mixtures at 5 K and identified the CN--containing species (e.g., CH$_2$CNH, CH$_3$NC, and CH$_2$NCH) upon X-ray irradiation of the C$_2$H$_2$$\cdots$NH$_3$ complex. Moreover, \cite{Canta2023} explored the vacuum ultraviolet (VUV) irradiation of C$_2$H$_2$ ices mixed with NH$_3$ and found that CH$_3$CHNH accounts for most of the depleted NH$_3$ as well as a small amount of CH$_3$CN.

In this work, we provide an experimental follow-up study on the interactions of C$_2$H$_2$ with NH$_2$ radicals and H atoms, which are formed upon VUV photodissociation of NH$_3$ ice, investigated at 10 K. In the interstellar ice, NH$_3$ is suggested to form through successive H--atom addition reactions to N atoms (i.e., N$\rightarrow$NH$\rightarrow$NH$_2$$\rightarrow$NH$_3$) along with H$_2$O in the translucent clouds \citep{Fedoseev2015a}. Therefore, the studied reactions can be triggered during the accretion of the atomic species N and H onto grain surfaces or by VUV photons penetrating the interstellar H$_2$O-rich ice containing NH$_3$ in molecular clouds. This work explores the relevant solid-state chemistry for the formation of N--bearing COMs on dust grains and reports on the possible chemical pathway.
The experimental details are described in the next section. The results and discussion are presented in Sections 3 and 4, respectively. Finally, Section 5 focuses on the astronomical relevance of the experimental data and provides our conclusions.

\section{Experiment}

All experiments were performed using the ultra-high vacuum (UHV) apparatus at the Laboratory Astrophysics and Cluster Physics in Jena. The details of the experimental setups and the latest characteristics of the applied UV-photon source (D$_2$ lamp) have been described in previous studies \citep{Potapov2019, Chuang2022}. Here, only the relevant information of this laboratory work is briefly presented. A KBr substrate is mounted on a copper sample holder connected to a closed-cycle helium cryostat through indium foil and located at the center of the UHV chamber. The base pressure is <1$\times$10$^{-9}$ mbar at room temperature. The substrate temperature was read out by silicon diodes with <0.5 K accuracy and regulated between 10 and 330 K through a Lakeshore temperature controller equipped with a resistive heater. A commercial D$_2$ lamp (Hamamatsu: D2H2--L11798), a MgF$_2$-sealed closed cycle light source, was used to generate VUV photons, which were further guided to the ice sample through another MgF$_2$ UHV-viewport installed on the chamber. The spectrum of the VUV photons is mainly characterized by broad emission centered at $\sim$160 nm, with negligible Ly-$\alpha$ at 121.6 nm due to the strong MgF$_2$ absorption at a lower wavelength. The UV flux of $\sim$7.6$\times$10$^{13}$ photon cm$^{-2}$s$^{-1}$ was estimated in \cite{Chuang2022} and agreed with values in the literature using a similar experimental setting \citep{Martin2020}. 

In this paragraph, we briefly describe the experimental procedure. Gaseous C$_2$H$_2$ (Air Liquide; 99.6\%) and NH$_3$ (Air Liquide, 99.999\%) were simultaneously introduced into the main chamber through two separate all-metal leak valves and deposited directly onto the pre-cooled substrate at 10 K. The deposition and the photolysis of ice samples were monitored in situ by Fourier-transform infrared spectroscopy (FTIR) in transmission mode in a range of 400-7500 cm$^{-1}$ with 1 cm$^{-1}$ resolution. The abundances of parent molecules and products in column density (\textit{N}, in molecule centimeter$^{-2}$) were obtained using the modified Beer-Lambert law \citep{Chuang2018thesis}. The absorption peak area was corrected by straight baseline subtraction and derived through Gaussian curve fitting, accompanied with one standard deviation as the error bars and/or direct integration of the IR feature. This estimation does not take into account the uncertainties due to the baseline subtraction and IR absorption band strength (\textit{A'}) of pure ices acquired from the literature. Theoretical band strength values were used for molecules such as vinylamine, acetaldeimine, ketenimine, and cyanide anion, which have no experimental value available. The applied band strength for the selected species is summarized in Table \ref{table01}. The ratios of the studied ice mixture are C$_2$H$_2$:NH$_3$=1:2.7, 1:1.4, 1:0.5, and 1:0.2, and the absolute abundance of the C$_2$H$_2$ ice is $\sim$(6-10)$\times$10$^{16}$ molecule cm$^{-2}$. These reported column densities and ratios can be recalibrated when more precise values become available.

After UV-photon irradiation of the studied ice sample, a temperature-programmed desorption (TPD) experiment with a ramping rate of 5 K min$^{-1}$ was performed. The quadrupole mass spectrometer recorded the sublimated species' mass signals as a function of temperature. The data of desorption temperature and the corresponding fragmentation pattern induced by the electron impact ionization at 70 eV were utilized to identify the newly formed products in addition to IR spectroscopy.


\section{Results}

\begin{figure}[]
        \begin{center}
                \includegraphics[width=90mm]{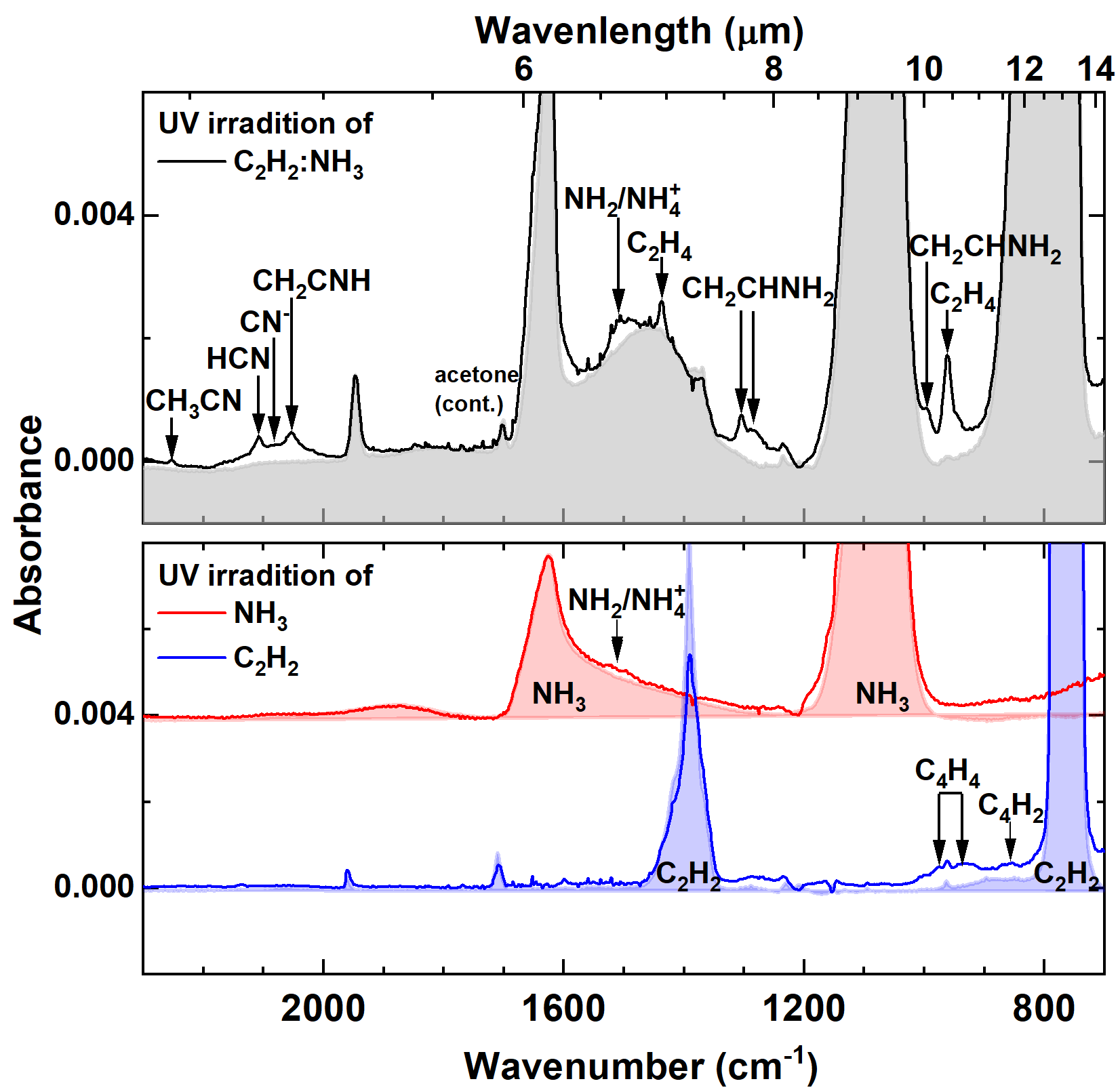}
                \caption{Infrared spectra of the studied interstellar ice analogs. Upper: IR spectra obtained after UV irradiation of the C$_2$H$_2$:NH$_3$ (1:1.4) ice mixture at 10 K for a fluence of 5.7$\times$10$^{17}$ photon cm$^{-2}$. The column densities of C$_2$H$_2$ and NH$_3$ in the ice mixture are 8.1$\times$10$^{16}$ and 1.1$\times$10$^{-17}$ molecule cm$^{-2}$, respectively. Bottom: Infrared spectra obtained after UV irradiation of pure ice NH$_3$ and C$_2$H$_2$ at 10 K for a fluence of 5.7$\times$10$^{17}$ photon cm$^{-2}$ and offset for clarity. The column densities of pure C$_2$H$_2$ and NH$_3$ ice are 7.3$\times$10$^{-17}$ and 1.1$\times$10$^{-17}$ molecule cm$^{-2}$, respectively. The shaded area is present for the IR spectra before UV irradiation. The IR peaks of the newly formed products are labeled with an arrow.}
                \label{Fig1}
        \end{center}
\end{figure}

Figure 1 presents the IR absorption spectra obtained after the UV irradiation of the C$_2$H$_2$:NH$_3$ (1:1.4) ice mixture (upper panel) and pure NH$_3$ as well as the C$_2$H$_2$ ice (bottom panel) at 10 K for a photon fluence of 5.7$\times$10$^{17}$ photon cm$^{-2}$. The shaded IR spectra obtained before the UV irradiation are presented for comparison; the newly formed IR features are shown in an unshaded area. The absorption peaks of the parent C$_2$H$_2$ are observed at $\sim$810 cm$^{-1}$ ($\nu$$_5$; CH bending), $\sim$1390 cm$^{-1}$ ($\nu$$_4$+$\nu$$_5$; combination), and 1950 cm$^{-1}$ (C$_2$H$_2$-relative band), and the peaks of NH$_3$ are found at $\sim$1067 cm$^{-1}$ ($\nu$$_2$; NH sym. deform) and 1625 ($\nu$$_4$; NH deg. deform) cm$^{-1}$. The tiny peak of acetone at 1710 cm$^{-1}$, a common contamination source in an acetylene gas bottle, is negligible; the ratio of \textit{N}(CH$_3$COCH$_3$) to \textit{N}(C$_2$H$_2$) is <0.1\%. It is noted that C$_2$H$_2$ IR features are different in pure and mixed ices with NH$_3$, indicating that the C$_2$H$_2$ spectral profiles (e.g., full width half maximum and peak position) are sensitive to the surrounding environment. The C$_2$H$_2$:NH$_3$ absorption spectra with different ratios are reported in the Appendix (Figure \ref{FigA1}) for future comparisons with experimental and observational data.

\begin{figure*}[t]
        \begin{center}
                \includegraphics[width=160mm]{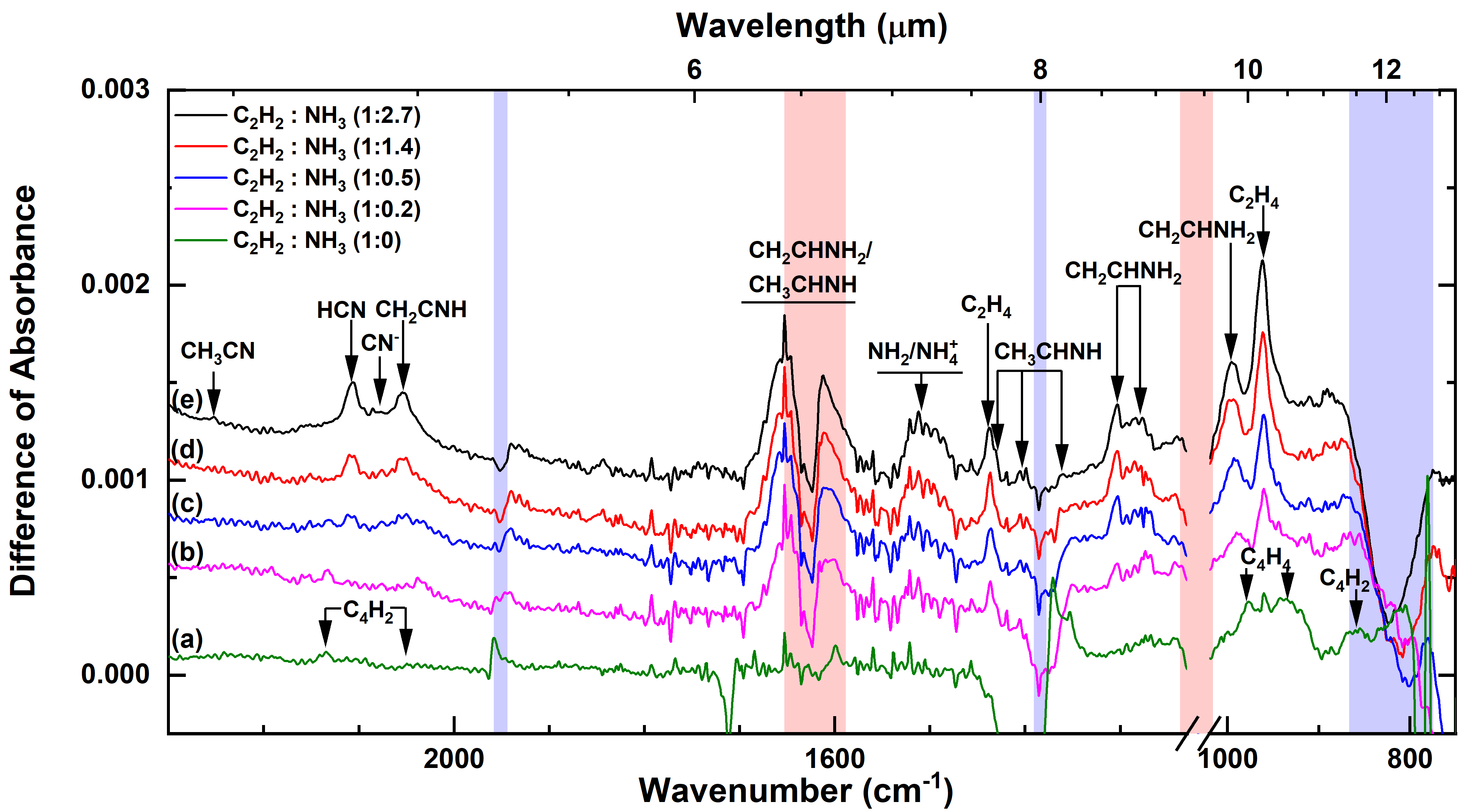}
                \caption{Difference IR spectra obtained after UV irradiation of C$_2$H$_2$:NH$_3$ ice mixtures with ratios of (a) 1:0, (b) 1:0.2, (c) 1:0.5, (d) 1:1.4, and (e) 1:2.7 at 10 K for a fluence of 5.7$\times$10$^{17}$ photon cm$^{-2}$. Infrared spectra are offset for clarity. The shaded area marks the absorption features of the parent species (i.e., blue is for C$_2$H$_2$ and red is for NH$_3$) now visible as negative bands.}
                \label{Fig2}
        \end{center}
\end{figure*}

In the upper panel of Fig. \ref{Fig1}, several IR peaks are exclusively detected after UV photolysis of the C$_2$H$_2$:NH$_3$ (1:1.4) ice mixture in addition to the absorption features of hydrocarbons (e.g., C$_2$H$_4$, C$_4$H$_2$, and C$_4$H$_4$) and NH$_2$/NH$_4^+$, which are also observed after UV irradiation of individual C$_2$H$_2$ and NH$_3$ ice, respectively (see the bottom panel). Therefore, these IR peaks must represent the newly formed products containing CN bonds. In the 2300-2000 cm$^{-1}$ region, where CN stretching modes are present, the absorption peak at 2253 cm$^{-1}$ has been assigned to acetonitrile CH$_3$CN ($\nu$$_2$). This is the strongest absorption feature, and its band strength is only 2.2$\times$10$^{-18}$ cm molecule$^{-1}$, which is an order of magnitude less than that of common interstellar molecules \citep{Hatch1992, Hudson2004, Hattori2005, Abdulgalil2012}. The rest of the CH$_3$CN IR features are most likely blended with other species due to their weak absorption coefficient. This detection is in agreement with \cite{Canta2023}, who have reported the formation of CH$_3$CN upon the VUV irradiation of C$_2$H$_2$:NH$_3$ = 1:1 ice mixtures. Furthermore, multiple overlapping peaks at 2107 cm$^{-1}$, 2081 cm$^{-1}$, and 2054 cm$^{-1}$ have been successfully deconvoluted using Gaussian curve fitting and assigned to hydrogen cyanide (HCN, $\nu$$_3$; CN stretching mode), cyanide anion (CN$^-$, $\nu$$_3$; CN stretching mode), and ketenimine (CH$_2$CNH, $\nu$$_3$; C=C=N stretching mode), respectively \citep{Jacox1979, Mungan1991, Ito2010, Gerakines2022}. The detailed IR identification of HCN and CN$^-$ refers to the early work by \cite{Gerakines2004}. Besides these (highly) unsaturated CN--bearing products (e.g., containing the moieties C-C$\equiv$N/C=C=N), the IR features of C$_2$H$_5$N isomers, for example, vinylamine (CH$_2$CHNH$_2$) and acetaldimine (CH$_3$CHNH), are found at 1302, 1283, 996 cm$^{-1}$ in Fig. \ref{Fig1}, while other characteristic peaks are blended with the parent NH$_3$ and C$_2$H$_2$ features. Acetaldimine has also been detected in previous VUV-irradiation works on C$_2$H$_2$:NH$_3$ ice mixtures \citep{Canta2023}. It is important to note that CH$_2$CHNH$_2$ and CH$_3$CHNH are tautomers (interconverted chemical structure; amine $\leftrightarrow$ imine).  
Aziridine (c-C$_2$H$_4$NH), one of the C$_2$H$_5$N isomers with the highest chemical potential energy (i.e., $\sim$49 and $\sim$43 kcal mol$^{-1}$ higher than CH$_3$CHNH and CH$_2$CHNH$_2$, respectively), is absent \citep{Stolkin1977}.

\begin{figure*}[t]
        \begin{center}
                \includegraphics[width=\textwidth]{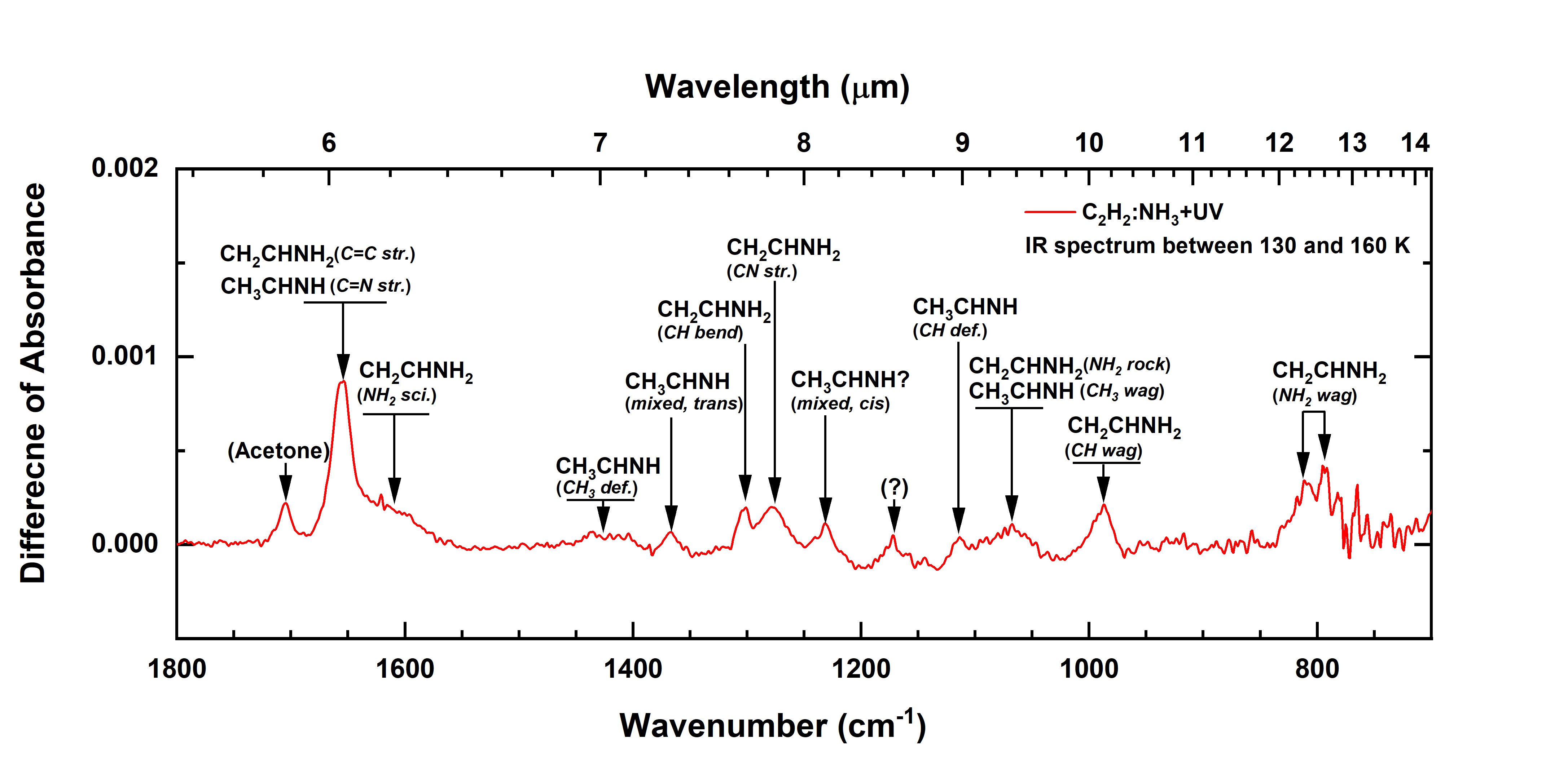}
                \caption{Infrared spectrum obtained at the interval of 130-160 K during the TPD experiment of the UV-irradiated ice mixture C$_2$H$_2$:NH$_3$ (1:1.4) at 10 K for a fluence of 5.7$\times$10$^{17}$ photon cm$^{-2}$.}
                \label{Fig3}
        \end{center}
\end{figure*}

The newly formed molecules are expected to result from the association of several intermediate species upon UV photolysis of the C$_2$H$_2$:NH$_3$ ice mixture. Therefore, the intensity of these designated IR peaks should be correlated with the concentration of C$_2$H$_2$ and/or NH$_3$. Figure \ref{Fig2} shows the IR difference spectra obtained after UV irradiation of C$_2$H$_2$:NH$_3$ ice mixtures with different ratios, including (a)1:0, (b) 1:0.2, (c) 1:0.5, (d) 1:1.4, and (e) 1:2.7 at 10 K for a fluence of 5.7$\times$10$^{17}$ photon cm$^{-2}$. The parent and product features appear as negative and positive bands in the difference spectra, respectively. For example, the negative peaks of C$_2$H$_2$ are found at $\sim$1950, $\sim$1390, and $\sim$810 cm$^{-1}$, and NH$_3$ peaks are shown at $\sim$1626 and $\sim$1067 cm$^{-1}$. In spectrum (a) of Fig 2, the detection of C$_4$H$_4$ ($\sim$1940, $\sim$977, and $\sim$935 cm$^{-1}$) and C$_4$H$_2$ ($\sim$2133, $\sim$2138, and $\sim$857 cm$^{-1}$) is in line with previous studies on photolysis (and radiolysis) of pure C$_2$H$_2$ \citep{Compagnini2009, Abplanalp2020, Lo2020, Pereira2020}. A tiny C$_2$H$_4$ peak can be seen at 961 cm$^{-1}$ and is due to a limited amount of H atoms for the successive H-addition reactions C$_2$H$_2$$\rightarrow$C$_2$H$_3$$\rightarrow$C$_2$H$_4$. However, in the spectra (b)-(e) of Fig. \ref{Fig2}, C$_2$H$_4$ IR features at 961 and 1438 cm$^{-1}$ become more dominant with the increasing concentration of NH$_3$ in C$_2$H$_2$:NH$_3$ ice mixtures. This is because the photodissociation of NH$_3$ mainly results in NH$_2$ ($\tilde{A}$ or $\tilde{X}$) radicals and H atoms \citep{Evans2012}. The newly generated H atoms enhance the C$_2$H$_2$ hydrogenation forming C$_2$H$_4$, while C$_2$H$_6$ remains absent. In contrast, C$_4$H$_4$ and C$_4$H$_2$ features become less dominant than other products. Besides the hydrocarbon formation, the photofragment NH$_2$ could react with C$_2$H$_2$ or its hydrogenated species, C$_2$H$_n$, forming N--bearing COMs. The IR spectral assignments of products containing CCN bonds are supported by observing their characteristic peaks, which increase with increasing NH$_3$ concentration. For example, in the spectrum (e) with the highest NH$_3$ concentration (C$_2$H$_2$:NH$_3$=1:2.7), the IR features of CH$_3$CN (2253 cm$^{-1}$), HCN (2107 cm$^{-1}$), CN$^-$ (2081 cm$^{-1}$), and CH$_2$CNH (2054 cm$^{-1}$) are the most abundant. The IR features of CH$_2$CHNH$_2$ at 1302, 1283, and 996 cm$^{-1}$ and of CH$_3$CHNH at 1367 cm$^{-1}$ are also enhanced, but their strongest peaks at $\sim$1660 cm$^{-1}$ overlap strongly with the negative feature of NH$_3$ at 1626 cm$^{-1}$. 

Given that the desorption temperatures of the newly formed N--bearing COMs are (much) higher than those of the parent C$_2$H$_2$ ($\sim$80 K) and NH$_3$ ($\sim$100 K), the corresponding IR features of CH$_2$CHNH$_2$ and CH$_3$CHNH can be revealed after the sublimation of the parent species. Figure \ref{Fig3} shows the IR difference spectrum obtained between 130 and 160 K, where CH$_2$CHNH$_2$ and CH$_3$CHNH are expected to desorb under the current experimental conditions. As pure ice spectra of these species are still lacking, the following assignment of vibrational modes mainly relies on theoretical studies and matrix isolation spectroscopy. In Fig. \ref{Fig3}, the strongest absorption feature is at $\sim$1660 cm$^{-1}$, which is due to vibrational transitions of CH$_2$CHNH$_2$ ($\nu$$_6$; C=C stretching mode) and CH$_3$CHNH ($\nu$$_5$; C=N stretching mode) \citep{Stolkin1977, Hamada1984, Hashiguchi1984, Lammertsma1994}. A broad feature at $\sim$1070 cm$^{-1}$ also has dual contributions from the NH$_2$ rocking mode ($\nu$$_{11}$) of CH$_2$CHNH$_2$ and the CH$_3$ wagging mode ($\nu$$_{10}$) of CH$_3$CHNH$_2$. Several isolated peaks of CH$_2$CHNH$_2$ were clearly detected at $\sim$1600 ($\nu$$_7$; NH$_2$ deformation mode), 1302 ($\nu$$_9$; CH bending mode), 1280 ($\nu$$_{10}$; CN stretching mode), 986 ($\nu$$_{12}$; CH waging mode), and $\sim$810 ($\nu$$_{14}$; NH$_2$ wagging) cm$^{-1}$ \citep{Hamada1984, Lammertsma1994, McNaughton1999}. The unblended IR features of CH$_3$CHNH are observed at 1435/1410 ($\nu$$_{14}$; CH$_3$ deformation modes), 1367/1231 ($\nu$$_8$; trans-/cis-mixed modes), and 1112 ($\nu$$_9$; CH deformation mode) cm$^{-1}$ \citep{Stolkin1977, Hashiguchi1984}. The presented IR spectrum at the interval of 130-160 K complementarily supports the previous N--bearing COM identification at 10 K, as shown in Fig. \ref{Fig1}. Moreover, it provides a temperature domain to ensure the spectral identification.

\begin{figure}[]
        \begin{center}
                \includegraphics[width=90mm]{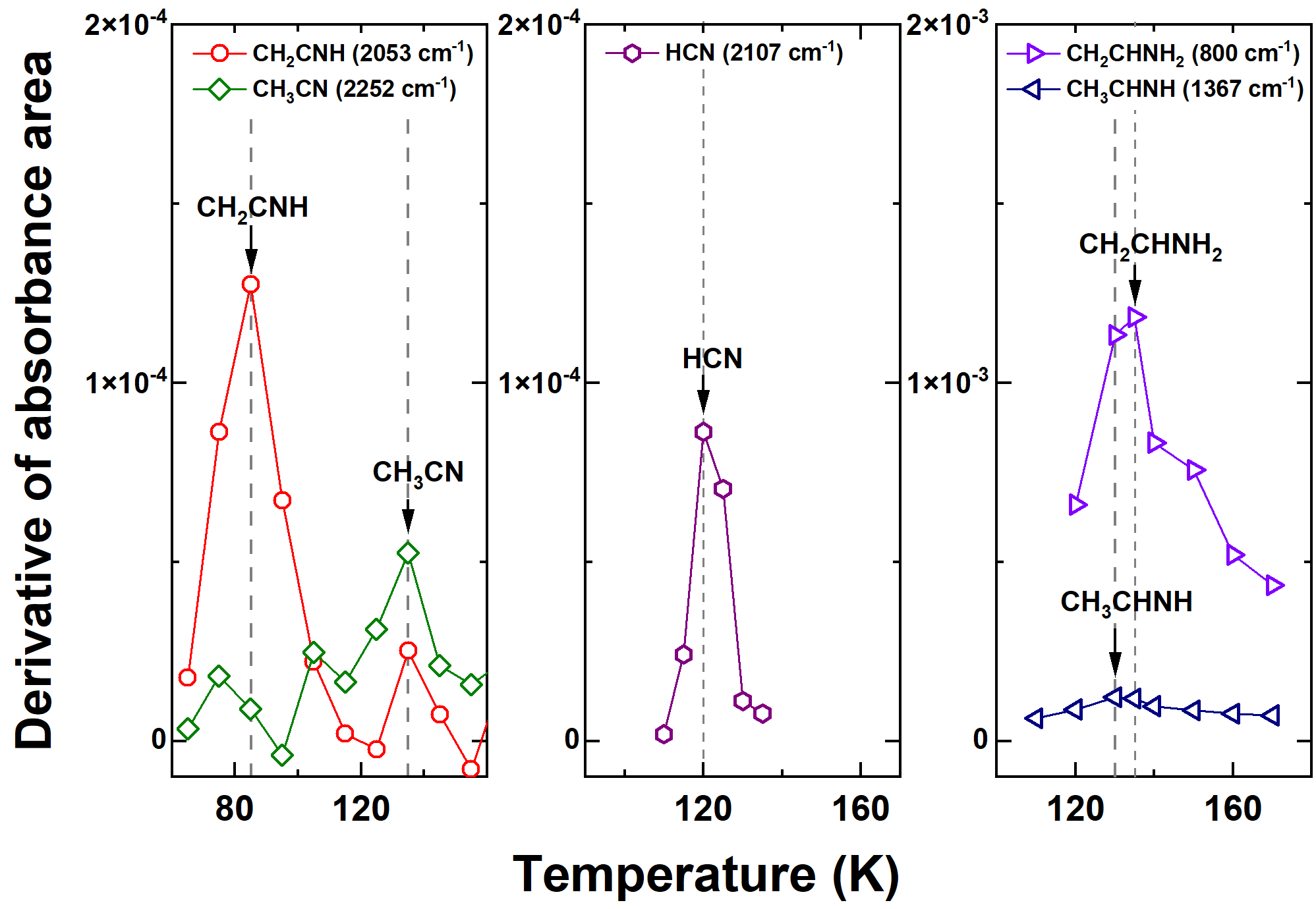}
                \caption{Derivatives of the species' absorbance area during the TPD experiment are presented as a function of temperature.}
                \label{Fig4}
        \end{center}
\end{figure}

Similarly, during the TPD experiment, the absorption peak intensities (i.e., absorbance area) of the assigned species, including C$_2$H$_4$ (960 cm$^{-1}$), HCN (2107 cm$^{-1}$), CH$_3$CN (2252 cm$^{-1}$), CH$_2$CNH (2053 cm$^{-1}$), CH$_3$CHNH (1367 cm$^{-1}$), and CH$_2$CHNH$_2$ ($\sim$810 cm$^{-1}$), were monitored as a function of the substrate temperature. Each obtained data point spans over $\sim$5 K due to the FTIR measurement time (i.e., 60 seconds). An overview of the IR absorbance area of the selected species during the TPD experiment is reported in the Appendix (see Figure \ref{FigA2}). Moreover, their derivatives of absorbance area as a function of temperature are reported in Fig. \ref{Fig4}, pointing to the desorption efficiency of the selected species. For example, in Fig. \ref{Fig4}, the C$_2$H$_3$N isomers, namely CH$_2$CNH and CH$_3$CN, desorb at quite different temperatures (i.e., $\sim$85 and $\sim$135 K, respectively; left panel), and HCN desorbs mainly at $\sim$120 K (middle panel). The desorption peaks of the C$_2$H$_5$N isomers, namely CH$_3$CHNH and CH$_2$CHNH$_2$, are at $\sim$130 K and $\sim$135 K, respectively (right panel). Due to the low resolution of the temperature readout for IR data, the desorption peaks of CH$_3$CHNH and CH$_2$CHNH$_2$ are challenging to pinpoint.  

\begin{figure}[b]
        \begin{center}
                \includegraphics[width=90mm]{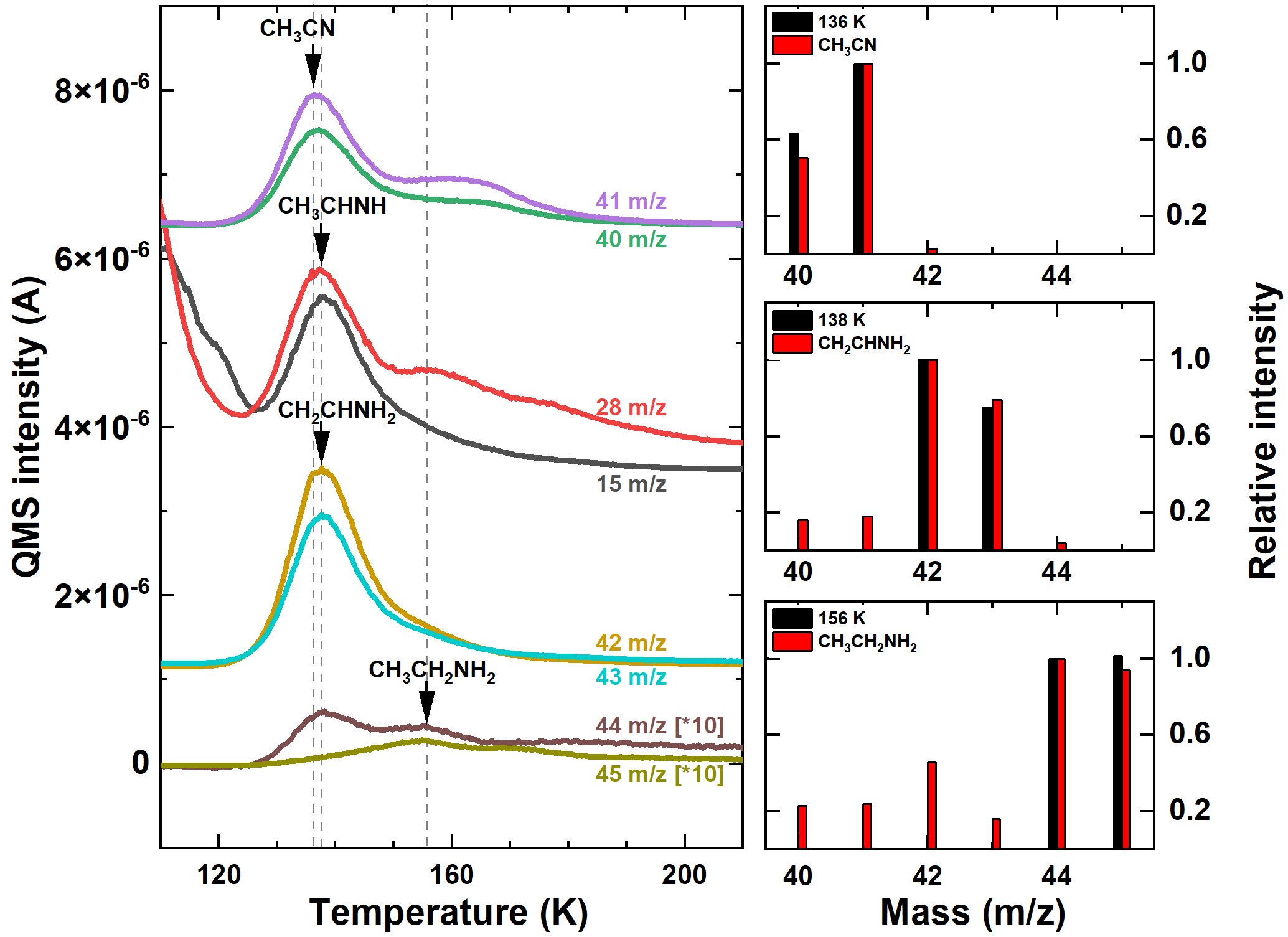}
                \caption{Species identification in QMS-TPD data. Left: Desorption signals of interested masses obtained during the TPD experiment after UV irradiation of C$_2$H$_2$:NH$_3$ (1:1.4) for a fluence of 5.7$\times$10$^{17}$ photon cm$^{-2}$ at 10 K. The ion signals have been shifted for clarity. The dashed lines indicate the desorption peak of the assigned molecules. Right: Comparison of the obtained mass fragmentation pattern at 136.4, 137.3/137.7, and 155.2 K with standard data for CH$_3$CN (NIST), CH$_2$CHNH$_2$ (SpectraBase), and CH$_3$CH$_2$NH$_2$ (NIST).}
                \label{Fig5}
        \end{center}
\end{figure}

\begin{figure}[b]
        \begin{center}
                \includegraphics[width=90mm]{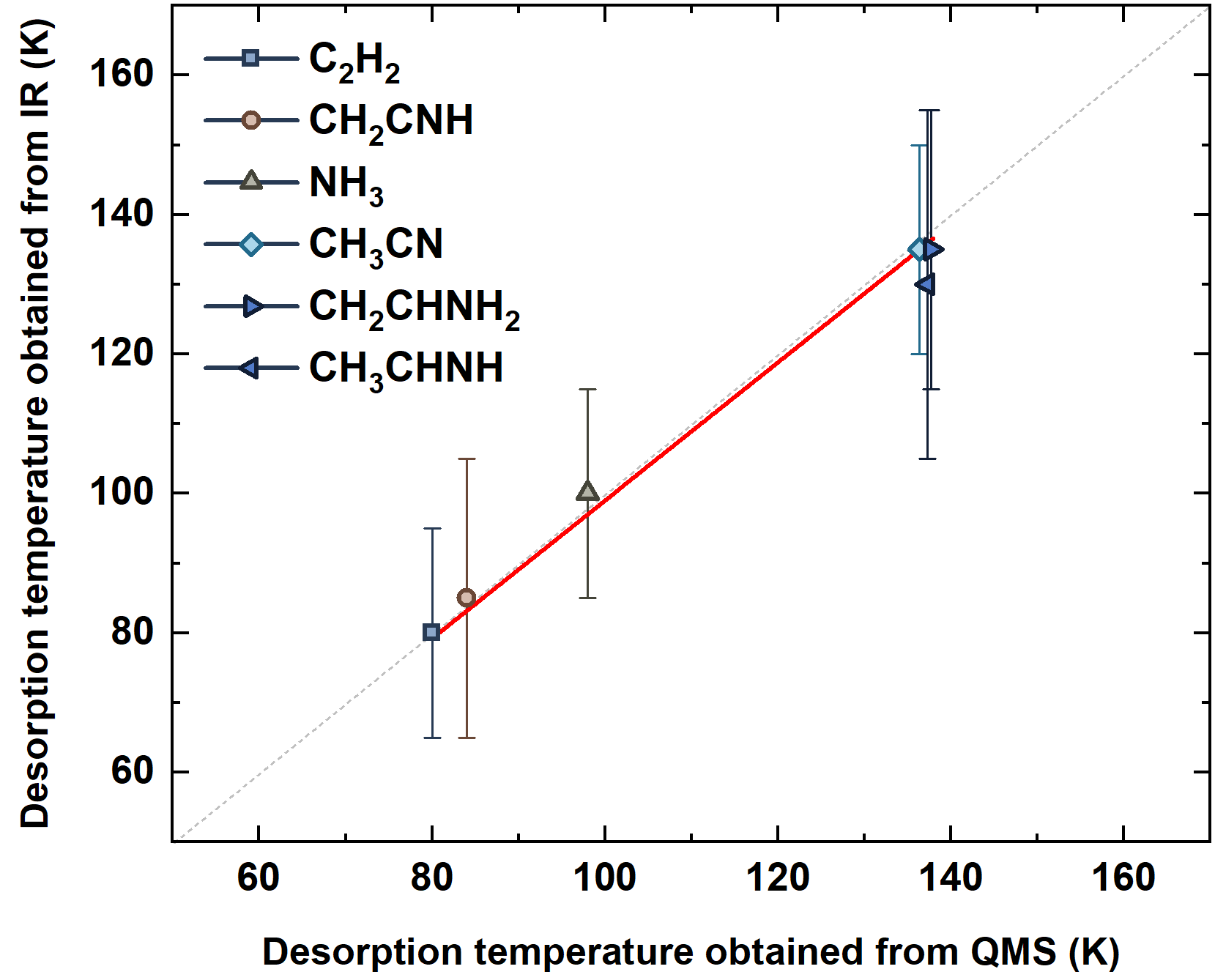}
                \caption{Desorption temperatures of selected species obtained from IR and QMS methods during the TPD experiment. The dashed line represents a one-to-one correlation, and the solid line (red) shows the linear fitting result. The vertical bar indicates the desorption range in the IR-TPD data.}
                \label{Fig6}
        \end{center}
\end{figure}

The left panel of Fig. \ref{Fig5} shows the desorption mass signals obtained by QMS during the TPD experiment after the UV irradiation of the ice mixture C$_2$H$_2$:NH$_3$ (1:1.4) for a fluence of 5.7$\times$10$^{17}$ photon cm$^{-2}$ at 10 K. Since QMS has a fast-scanning cycle (approximately every five seconds), each data point only spans over $\sim$0.4 K, giving a better temperature resolution. The prominent desorption peaks of CH$_3$CN, for example, 40 (C$_2$H$_2$N$^+$) and 41 (C$_2$H$_3$N$^+$) m/z, are clearly shown at $\sim$136.4 K, which is in line with the IR data. The desorption peaks of 15 (CH$_3$$^+$/NH$^+$) and 28 (CH$_2$N$^+$) m/z centered at $\sim$137.3 K as well as 42 (C$_2$H$_4$N$^+$) and 43 (C$_2$H$_5$N$^+$) m/z centered at $\sim$137.7 K are most likely from two C$_2$H$_5$N isomers. Although the desorption temperature of the above mass signals is very close to CH$_3$CN, the selected mass signals are exclusively from C$_2$H$_5$N isomers due to the higher molecular masses. At higher temperatures, we found two main desorption mass signals of 44 (C$_2$H$_6$N$^+$) and 45 (C$_2$H$_7$N$^+$) m/z at 155.2 K, which can be attributed to CH$_3$CH$_2$NH$_2$. In the right panel of Fig. \ref{Fig5}, the identification of QMS-TPD data is carefully confirmed by comparing these molecular-relevant mass signals with the species’ standard fragmentation pattern induced by electron impact. The data of CH$_3$CN and CH$_3$CH$_2$NH$_2$ are from the NIST database,\footnote{https://doi.org/10.18434/T4D303} and CH$_2$CHNH$_2$ is from the SpectralBase.\footnote{https://spectrabase.com/} It is important to note that the reported CH$_2$CHNH$_2$ fragment spectrum has no visible contribution from 28 and 15 m/z mass signals. However, \cite{Vinogradoff2012} showed that the primary fragment mass of CH$_3$CHNH is 28 m/z (CH$_2$N$^+$) due to the loss of the CH$_3$ fragment (15 m/z), while mass signals of 42 and 43 m/z only account for <10\%. Therefore, the 28 and 15 m/z mass signals detected at $\sim$137.3 K are expected to originate primarily from the CH$_3$CHNH. The desorption temperatures obtained from IR-TPD and QMS-TPD data, including parent and product species, are summarized in Fig. \ref{Fig6}. A linear fit to the data set shows an excellent agreement (i.e., the fit slope of $\sim$0.99$\pm$0.01) between two independent detection methods, complementarily identifying these N--bearing products. 

\section{Discussion}

\begin{figure}[t]
        \begin{center}
                \includegraphics[width=85mm]{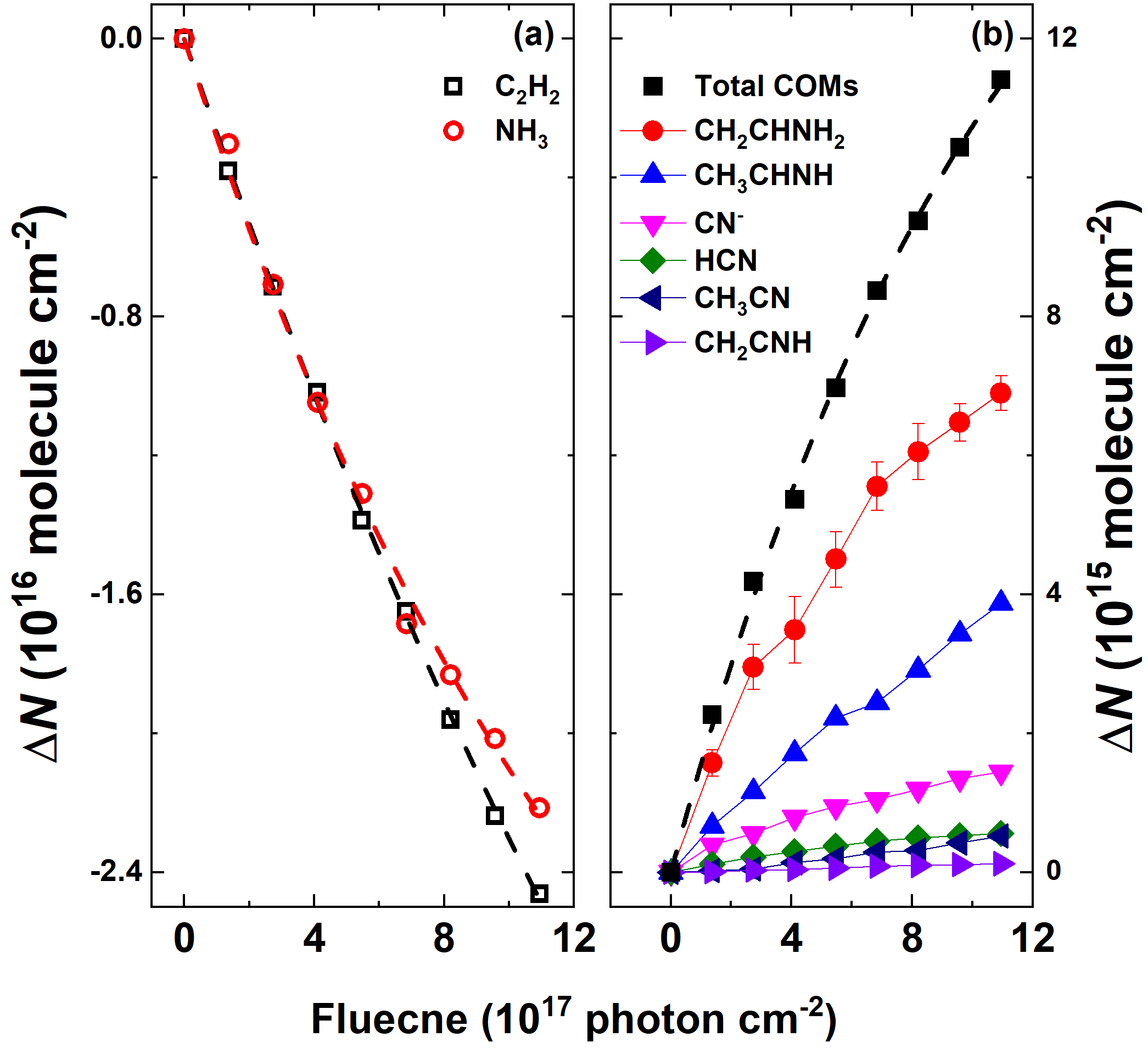}
                \caption{Kinetic evolution of (a) parent and (b) products derived from UV photolysis of the C$_2$H$_2$:NH$_3$ (1:1.4) ice mixture at 10 K over a fluence of 1.1$\times$10$^{18}$ photon cm$^{-2}$. The dashed lines present the fitting results, and the solid lines connecting data are only for clarity.}
                \label{Fig7}
        \end{center}
\end{figure}

Figure \ref{Fig7} shows the temporal evolution of the parent and newly formed products obtained in the UV photolysis of a C$_2$H$_2$:NH$_3$ (1:1.4) ice mixture at 10 K over a fluence of 1.1$\times$10$^{18}$ photon cm$^{-2}$. The molecular abundances are derived from their corresponding IR absorption features. The obtained products' kinetic evolution in the UV-photolyzed C$_2$H$_2$:NH$_3$ ice mixture shows possible formation routes. 

In the left panel of Fig. \ref{Fig7}, the depletion of C$_2$H$_2$ and NH$_3$ are reported and fit with a single exponential equation:
\begin{equation}
\label{Eq01}
 \Delta~\textit{N}\text{(molecules)} = \alpha(1 -\text{exp}(\sigma~\cdot~\phi~\cdot~t)),
\end{equation}
where $\alpha$ is the saturation value (i.e., the maximum abundance when reaching the equilibrium state) in molecule centimeter$^{-2}$, \textit{$\sigma$} is the effective cross-section in centimeter$^2$, $\phi$ is the UV flux in photon centimeter$^{-2}$ s$^{-1}$, and \textit{t} is the irradiation time in seconds. The derived photolysis cross-section values are (3.7$\pm$0.4)$\times$10$^{-19}$ and (7.1$\pm$1.2)$\times$10$^{-19}$ cm$^2$ for C$_2$H$_2$ and NH$_3$, respectively. 
Given that the applied UV photons dominate at 130-160 nm (i.e., photon energy of 7.7-9.5 eV) with a negligible amount of Ly-$\alpha$ photons, the direct photoionization of C$_2$H$_2$ and NH$_3$ is excluded under the current experimental conditions due to their relatively high threshold energies (i.e., $\sim$11.4 and $\sim$10.2 eV for C$_2$H$_2$ and NH$_3$, respectively; \citep{Collin1967, Xia1991, Locht1991}). Moreover, the C$_2$H$_2$ photodissociation primarily results in the formation of an ethynyl radical (CCH) instead of forming two methylidyne radicals (CH) because of a strong C$\equiv$C bond requiring a threshold energy of 9.9 eV (photon wavelength $\sim$125.3 nm; \citealt{Okabe1975}) for the dissociation. This explains the main product formation with an even number of carbon atoms, such as C$_4$H$_2$, through recombining two C$_2$H radicals and the absence of CH$_4$ in the pure C$_2$H$_2$ ice experiment. The depletion mechanism of NH$_3$ is most likely due to photodissociation resulting in NH$_2$ and H \citep{Okabe1967};
 \begin{equation}
\label{Eq02}
\text{NH$_3$} \xrightarrow{{+h\nu}} \text{NH$_2$ + H}.
\end{equation}
As mentioned previously, the detected products result from interactions among photofragments of the C$_2$H$_2$:NH$_3$ ice mixture. For example, H reacts with C$_2$H$_2$, leading to the formation of C$_2$H$_3$ through H--atom addition reactions. The C$_2$H$_2$ hydrogenation on dust grains at 10-20 K has been investigated in theoretical and laboratory studies, and their activation energies are overcome by quantum tunneling at low temperatures \citep{Hiraoka2000, Kobayashi2017};
\begin{equation}
\label{Eq03}
\text{C$_2$H$_2$+H} \rightarrow\text{C$_2$H$_3$}.
\end{equation}
The estimated reaction barrier of reaction \ref{Eq03} (4.30-4.78 kcal mol$^{-1}$) is very similar to the value of the reaction CO+H $\rightarrow$ HCO (3.25 kcal mol$^{-1}$), which is one of the critical surface reactions on interstellar dust grains \citep{Miller2004, Kobayashi2017, Alvarez2018}. The efficient conversion from triple-bond (C$\equiv$C) into double-bond (C=C) species is expected upon H--atom addition under molecular cloud conditions.

The radical-molecule association between NH$_2$ and C$_2$H$_2$ has been experimentally reported in the gas phase and further supported by theoretical calculation showing a relatively high activation barrier of 6.93-8.24 kcal mol$^{-1}$ \citep{Bosco1984, Lesclaux1985, Hennig1995, Moskaleva1998};
\begin{equation}
\label{Eq04}
\text{C$_2$H$_2$+NH$_2$} \rightarrow\text{CHCHNH$_2$}.
\end{equation}
Moreover, it has been pointed out that the reaction rates of reaction \ref{Eq04} in the gas phase and solid state are several orders of magnitude lower than that of C$_2$H$_2$+H and C$_2$H$_2$+OH \citep{Brunning1985, Molpeceres2022}. Therefore, reaction \ref{Eq04} is a critical step, and whether this route plays a relevant role in interstellar ice chemistry probably depends on the excess energy or excited state of NH$_2$ after NH$_3$ photodissociation. In addition to a lower activation barrier for reaction \ref{Eq03}, H atoms are more mobile than NH$_2$ at 10 K, which gives them a higher chance to meet C$_2$H$_2$ in the interstellar ice. Consequently, the competition between reactions \ref{Eq03} and \ref{Eq04} favors the C$_2$H$_3$ formation. Since the radical-radical associations are generally considered barrierless reactions, the recombination between C$_2$H$_3$ and nearby NH$_2$ could lead to the CH$_2$CHNH$_2$ formation through the reaction
\begin{equation}
\label{Eq05}
\text{C$_2$H$_3$ + NH$_2$} \rightarrow\text{CH$_2$CHNH$_2$}.
\end{equation}
This reaction route has been reported in gas-phase laboratory studies and used to explain the CN--bearing species in planetary chemical models \citep{Ferris1988, Kaye1983}. Very recently, \cite{Zhang2023} also reported this reaction route forming CH$_2$CHNH$_2$ in their laboratory work studying the 5--keV electron bombardment of a C$_2$H$_2$:NH$_3$ ice mixture. In Fig. \ref{Fig7} (b), CH$_2$CHNH$_2$ is immediately and abundantly formed upon the impact of UV photons, suggesting that CH$_2$CHNH$_2$ is a first-generation product in the C$_2$H$_2$:NH$_3$ ice mixture. Furthermore, its tautomer CH$_3$CHNH can also be observed and has a nearly linear formation. The ratio of CH$_3$CHNH/CH$_2$CHNH$_2$ increases with UV fluence. A similar formation curve has been found for the formation of keto-enol tautomers, namely CH$_2$CHOH and CH$_3$CHO, in the previous study of H$^+$ radiolysis of a C$_2$H$_2$:H$_2$O ice mixture at 17 K (see Figure 4 in \citealt{Chuang2021}).

The keto--enol tautomerization is expected to proceed through intermolecular H relocations enhanced by H$_2$O \citep{Chuang2020}. The amine--imine tautomerization of C$_2$H$_5$N isomers has been theoretically investigated and suggests the preference of the imine (acetaldimine) over the amine (vinylaminy) form. The potential energy of CH$_3$CHNH is $\sim$4 kcal mol$^{-1}$ lower than that of CH$_2$CHNH$_2$ \citep{Lammertsma1994}. However, the activation energy via intramolecular displacement is around 66.43-84.52 kcal mol$^{-1}$ in the gas phase, which is similar to the intramolecular shift barrier from CH$_2$CHOH to CH$_3$CHO (see Table 4 in \citealt{Apeloig1990, Lin1995, Andres1998}). It has been suggested that both interconversions require a catalyst \citep{Raczynska2005}. In this work, the simultaneous detection of CH$_3$CHNH and CH$_2$CHNH$_2$ hints at a possible tautomerization taking place in the studied chemical system as explained for the enol-keto species; 
\begin{equation}
\label{Eq06}
\text{CH$_2$CHNH$_2$ (amine-form)} \leftrightarrow\text{CH$_3$CHNH (imine-form)}.
\end{equation}

The consequent photodissociation of C$_2$H$_5$N isomers has been reported to form HCN in the gas phase \citep{Kaye1983}. In Fig. \ref{Fig7} (b), the kinetic evolution of HCN and its more stable ionic format, CN$^-$ are correlated. That is because that part of the HCN is expected to be converted into CN$^-$ through the acid-base reactions with NH$_3$ \citep{Noble2013}:
\begin{equation}
\label{Eq07}
\text{HCN+NH$_3$} \leftrightarrow\text{[NH$_4^+$][CN$^-$]}.
\end{equation}
The yields of HCN and CN$^-$ are shown upon the formation of C$_2$H$_5$N isomers; the derived product ratios of CH$_2$CHNH$_2$/HCN=8.35$\pm$0.36 and CN$^-$/HCN=5.39$\pm$0.38 are relatively constant over the entire UV fluence. These results indicate the efficient UV photodissociation of the first-generation products (H$_2$CCHNH$_2$ or CH$_3$CHNH) leading to HCN and [NH$_4^+$][CN$^-$]. 

The detection of unsaturated C$_2$H$_3$N molecules (e.g., CH$_2$CNH and CH$_3$CN) at a later irradiation stage is probably caused by the direct H$_2$ (or 2H) elimination induced by UV photons:
\begin{equation}
\label{Eq08}
\text{CH$_2$CHNH$_2$} \xrightarrow{+h\nu}\text{CH$_2$CNH + H$_2$ (or 2H)}.
\end{equation}
and
\begin{equation}
\label{Eq09}
\text{CH$_3$CHNH} \xrightarrow{+h\nu}\text{CH$_2$CNH / CH$_3$CN + H$_2$ (or 2H)}.
\end{equation}
The obtained abundance \textit{N}(CH$_3$CN) is about four times higher than the \textit{N}(CH$_2$CNH), which aligns with their order in potential energy (CH$_2$CNH is $\sim$26.1 kcal mol$^{-1}$ higher than CH$_3$CN; \citealt{Dickerson2018}). In addition to direct UV dissociation, other possible mechanisms might affect the ratio of CH$_3$CN and CH$_2$CNH. The isomerization from CH$_3$CN to CH$_2$CNH and CH$_3$NC with high activation barriers of $\sim$99.5 and $\sim$64.4 kcal mol$^{-1}$, respectively, have been found to be caused only by proton, X-ray, or N-atom bombardment, but investigation of this is outside the scope of this work \citep{Hudson2004, Mencos2016, Kameneva2017}. 

Apart from UV photodissociation, the newly formed CH$_2$CHNH$_2$ or CH$_3$CHNH are expected to react with H atoms, forming CH$_3$CH$_2$NH$_2$:
\begin{equation}
\label{Eq10}
\text{CH$_2$CHNH$_2$ (or CH$_3$CHNH) + 2H } \rightarrow\text{CH$_3$CH$_2$NH$_2$}.
\end{equation}
However, this analog reaction is speculated based on the H--atom addition reactions to vinyl alcohol (or acetaldehyde) forming ethanol. The derived QMS data suggest the CH$_3$CH$_2$NH$_2$ formation and support a general chemical conversion in the interstellar ice from H--poor into H--rich species \citep{Chuang2021}. It is important to note that a direct laboratory study of the H-addition and H-abstraction reactions linking C$_2$H$_n$N (n=3, 5, and 7) is still desired. 

The total abundance of N-bearing COMs, including CH$_2$CHNH$_2$, CH$_3$CHNH, CH$_3$CN, and CH$_2$CNH, is presented in the right panel of Fig. \ref{Fig7} and fit with a single exponential equation (Eq. \ref{Eq01}). The CH$_3$CH$_2$NH$_2$ yield is below the current IR detection sensitivity. Therefore, it is not counted in the total COM abundance. The derived effective formation cross-section for the total abundance of N-bearing COMs is (9.0$\pm$0.7)$\times$10$^{-19}$ cm$^2$. The sound pseudo-first-order fit suggests that these N-bearing COMs are formed through the common precursor C$_2$H$_2$. The conversion ratio of the total abundance of COMs to depleted C$_2$H$_2$, that is, \textit{N}(COMs)/$\Delta$\textit{N}(C$_2$H$_2$), changes with UV fluence, varying from 0.60 to 0.46. The decrease in the ratio is probably due to the further UV photolysis of N-bearing COMs.


\section{Astrochemical implication and conclusions}

In this work, we extended our previous studies on the reactions between acetylene and OH/H, which results in the formation of O--bearing COMs, to reactions between C$_2$H$_2$ and NH$_2$/H. This study provides experimental evidence on the solid-state formation of N--bearing COMs described by the formula C$_2$H$_n$N, including vinylamine (CH$_2$CHNH$_2$), acetaldimine (CH$_3$CHNH), acetonitrile (CH$_3$CN), and ketenimine (CH$_2$CNH), in the interstellar C$_2$H$_2$:NH$_3$ ice analog upon UV irradiation at 10 K. The identification of the above COMs was realized by IR spectroscopy and mass spectrometry. The correlation of the desorption temperatures of the species obtained from IR and QMS data during the TPD experiment complements the assignments. The initial molecules and newly formed products were monitored in situ as a function of the UV fluence, suggesting that CH$_2$CHNH$_2$ is the first-generation product that might tautomerize into CH$_3$CHNH, as found for the analog O--bearing species (e.g., CH$_2$CHOH and CH$_3$CHO) in the C$_2$H$_2$:H$_2$O ice mixture. The chemical derivatives with different degrees of hydrogenation, namely CH$_3$CN/CH$_2$CNH and CH$_3$CH$_2$NH$_2$, are formed through photolysis and hydrogenation of C$_2$H$_5$N isomers, respectively. The studied ice chemistry is of astronomical relevance, as it aids in explaining the complexity of N-bearing molecules in early star-forming stages and draws out the chemical relation among these observed species.

Gaseous C$_2$H$_2$ is commonly detected in the circumstellar shell of carbon-rich AGB stars such as IRC+10216 and several star-forming regions. Its abundance is about a few percent of H$_2$O: 1-5\% in massive young stellar objects, $\sim$1\% in protoplanetary disks, and 0.2-0.9\% in cometary comae \citep{Lacy1989, Brooke1996, Cernicharo1999, Lahuis2000, Mumma2003, Carr2008}. Photofragments of ionized polycyclic aromatic hydrocarbons (PAHs) forming C$_2$H$_2$ and C$_2$H by hard UV photons have been theoretically investigated and confirmed in the laboratory \citep{Jochims1994, Allain1996, Kress2010, Zhen2014, West2018}. Very recently, the JWST-MINDS team reported the abundant detection of C$_2$H$_2$ in a very low-mass star, 2MASS-J16053215-1933159, along with other complex hydrocarbons, C$_4$H$_2$ and C$_6$H$_6$ \citep{Tabone2023, Kamp2023}. The high abundance has been explained by the efficient (thermal) destruction of carbonaceous materials leading to the formation of C$_2$H$_2$ (\citealt{Anderson2017, vanDishoeck2023} and reference therein). In addition, C$_2$H$_2$ can be formed in the gas phase through dissociative electron recombination or proton transfer of C$_2$H$_3^+$. So far, however, the detection of C$_2$H$_2$ ice is limited to one observational study toward Serpens BG1 (see details in \citealt{Knez2008}), suggesting an abundance of $\sim$3\% with respect to H$_2$O ice. The lack of (more) solid detections of C$_2$H$_2$ ice has been explained by a difficult spectral identification. For example, the surrounding environment can significantly affect its absorption profile, and this is in addition to a severe blending with strong silicates and H$_2$O features at $\sim$13.5 $\mu$m \citep{Boudin1998, Knez2012}. 

In translucent clouds, atomic species accrete and associate on grain surfaces, resulting in an H$_2$O-rich ice layer containing CO$_2$, NH$_3$, and CH$_4$. For example, successive H--atom addition reactions to oxygen, nitrogen, and carbon consequently lead to H$_2$O, NH$_3$, and CH$_4$ ice formation, respectively (see review paper \citealt{Hama2013, Linnartz2015}). {The solid-state reactions involving the simplest alkyne C$_2$H$_2$ with atomic species (H and O atoms) have been proposed in theory to form various O-bearing COMs through the intermediate product C$_2$H$_3$ (\citealt{Tielens1992}, see also Figure 2 in \citealt{Charnley2004}). Furthermore, \cite{Charnley2001} has suggested that C$_2$H$_3$ could also react with N and H atoms to form several N-bearing cyclic species, including 1H-azirine, 2H-azirine, and aziridine, in addition to the reported molecules in this work. Recently, the interactions of C$_2$H$_2$ with simple radicals (e.g., OH, NH$_2$, and CH$_3$) have been theoretically studied to report their reaction rate constants in the solid state \citep{Molpeceres2022}. In the laboratory,} the hydrogenation of C$_2$H$_2$ ice has been investigated, showing the formation of (semi-)saturated hydrocarbons, such as C$_2$H$_4$ and C$_2$H$_6$ \citep{Hiraoka2000, Kobayashi2017}. Also, C$_2$H$_2$ ice can react with OH radicals, an important intermediate species in forming H$_2$O or a common fragment upon energetic processing of H$_2$O. In our previous studies, several O--bearing COMs, including vinyl alcohol, acetaldehyde, and ethanol, were detected in reactions between C$_2$H$_2$ and H/OH, which were formed through atomic association or proton radiolysis of H$_2$O ice \citep{Chuang2020, Chuang2021}. These results also provide one of the explanations for the observational discrepancy of COM compositions between early and late star-forming stages. The present experimental work further investigated possible reactions of C$_2$H$_2$ with NH$_2$ radicals and H atoms produced in the interstellar bulk ice upon impact of UV photons on a C$_2$H$_2$:NH$_3$ ice mixture. Given the higher activation barrier (i.e., $\sim$7.6 kcal mol$^{-1}$) required for the reaction C$_2$H$_2$+NH$_2$$\rightarrow$C$_2$H$_2$NH$_2$ 
reported by \cite{Molpeceres2022}, the formation of N--bearing COMs in the solid state most likely takes place through the recombination between NH$_2$ and C$_2$H$_3$ radicals. The latter are efficiently formed through the reaction C$_2$H$_2$+H$\rightarrow$C$_2$H$_3$ with a rate constant of $\sim$10$^{4}$ s$^{-1}$ \citep{Kobayashi2017, Molpeceres2022}. 

{Although the external interstellar radiation field is shielded by dust grains in clouds, a non-negligible UV flux can be induced by direct cosmic ray impacts on H$_2$ molecules through electronic excitations. The subsequent electron relaxation from the excited states results in the emissions of Lyman and Werner band photons \citep{Prasad1983}. The calculated cosmic ray-induced secondary UV flux varies between $\sim$10$^3$ and $\sim$3$\times$10$^4$ photon cm$^{-2}$s$^{-1}$, depending on the CR flux and R$_v$ (i.e., dust extinction curve in the optical region)\citep{Prasad1983, Cecchi1992, Shen2004}. A recent paper by \cite{Padovani2024} recalculated the secondary UV emissions using various CR ionization rates and Rv values and reported the photon flux in molecular clouds as a function of H$_2$ column density (see Figure 11 in the paper). Given the typical molecular cloud lifetime of $\sim$10$^{7}$ yr before ice sublimation due to the central star \citep{Chevance2020}, the total UV photons accumulated in clouds are expected to be in the range of 3$\times$10$^{17}$-1$\times$10$^{19}$ photon cm$^{-2}$. Therefore, the studied photochemistry over UV fluence of $\sim$1.1$\times$10$^{18}$ photon cm$^{-2}$ in this work is likely to take place within the lifetime of molecular clouds.} 
This work reports on the possible interstellar (bulk) ice chemistry triggered by secondary UV photons to form N--bearing species in molecular clouds. Furthermore, a recent astronomical observation of CH$_2$CHNH$_2$, CH$_3$CHNH, and (tentatively) CH$_3$CH$_2$NH$_2$ has been reported for the first time toward a quiescent giant molecular cloud, G+0.693--0.027 (\citealt{Zeng2021}, Rivilla et al., private communication). The simultaneous detections of C$_2$H$_n$N species suggest that they may share the same formation history and are chemically connected. Moreover, the astronomically observed CH$_2$CHNH$_2$ abundance is higher than that of CH$_3$CH$_2$NH$_2$, that is, $\textit{N}$(CH$_2$CHNH$_2$)/$\textit{N}$(CH$_3$CH$_2$NH$_2$) =1.7$\pm$0.5, which is consistent with the laboratory results showing the abundant formation of CH$_2$CHNH$_2$ over other species. 

The experimental findings from this study of the UV irradiation of C$_2$H$_2$:NH$_3$ ice mixtures at 10 K are summarized below:

\begin{enumerate}
    \item The interactions of C$_2$H$_2$ with NH$_2$ radicals and H atoms, which are produced by UV photolysis of NH$_3$, lead to the formation of several N--bearing complex molecules described by the formula C$_2$H$_n$N, such as vinylamine (CH$_2$CHNH$_2$), acetaldimine (CH$_3$CHNH), acetonitrile (CH$_3$CN), ketenimine (CH$_2$CNH), and ethylamine (CH$_3$CH$_2$NH$_2$), in interstellar ice analogs at 10 K.
    \item The simultaneous detection of CH$_2$CHNH$_2$ and CH$_3$CHNH indicates that isomerization takes place in the studied chemical system. It is speculated that intermolecular H--atom exchange with surrounding H---bearing species occurs as proposed previously for the O--bearing analogs, namely CH$_2$CHOH and CH$_3$CHO, in the C$_2$H$_2$:H$_2$O ice mixture.
    \item The kinetic evolution of these newly formed products shows that CH$_2$CHNH$_2$ is the first-generation product in the UV photolyzed C$_2$H$_2$:NH$_3$ ice mixture at 10 K, followed by the formation of CH$_3$CHNH. Both can be converted into chemically relevant species through direct dissociation and hydrogenation forming CH$_2$CNH/CH$_3$CN and CH$_3$CH$_2$NH$_2$, respectively.

\end{enumerate}

This present work verifies and extends our previous studies on the solid-state chemistry of the unsaturated hydrocarbon C$_2$H$_2$ as a prevalent formation pathway toward COMs by reactions with different functional groups, including astronomically relevant OH and NH$_2$ species, under molecular cloud conditions. It is important to note that fast H--atom addition reactions to C$_2$H$_2$ forming C$_2$H$_3$, thanks to quantum tunneling at low temperatures, could provide an alternative barrierless pathway to allow for association with some radicals that require a high activation energy to attach to C$_2$H$_2$. Following the studies on O-- and N--bearing COMs, possible reactions of C$_2$H$_2$ with SH radicals and H atoms forming S--bearing COMs are expected and currently under investigation (Santos et al., in preparation). The proposed chemistry of the simplest alkyne, a common product of the so-called top-down mechanism based on PAHs and hydrogenated amorphous carbonaceous grains, provides an efficient reaction network forming complex organics in interstellar environments. The contribution of different (non-)energetic processing will need further laboratory work and astrochemical models.

\begin{acknowledgements}
K.J.C. is grateful for support from the Dutch Research Council (NWO) via a VENI fellowship (VI.Veni.212.296). J.C.S. thanks the Danish National Research Foundation for the support through the Center of Excellence “InterCat” (DNRF150). Th.H. acknowledges support from the European Research Council (ERC) under the Horizon 2020 Framework Programme via the ERC Advanced Grant Origins 83 24 28. 
\end{acknowledgements}



\bibliography{Ref}{}

\begin{thebibliography}{}
\expandafter\ifx\csname natexlab\endcsname\relax\def\natexlab#1{#1}\fi
\providecommand{\url}[1]{\href{#1}{#1}}
\providecommand{\dodoi}[1]{doi:~\href{http://doi.org/#1}{\nolinkurl{#1}}}
\providecommand{\doeprint}[1]{\href{http://ascl.net/#1}{\nolinkurl{http://ascl.net/#1}}}
\providecommand{\doarXiv}[1]{\href{https://arxiv.org/abs/#1}{\nolinkurl{https://arxiv.org/abs/#1}}}

\bibitem[{Abdulgalil {et~al.}(2012)Abdulgalil, Marchione, Rosu-Finsen,
  Collings, McCoustra, {et~al.}}]{Abdulgalil2012}
Abdulgalil, A.~G., Marchione, D., Rosu-Finsen, A., {et~al.} 2012, Journal of
  Vacuum Science \& Technology A, 30

\bibitem[{Abplanalp \& Kaiser(2020)}]{Abplanalp2020}
Abplanalp, M.~J., \& Kaiser, R.~I. 2020, ApJ, 889, 3

\bibitem[{Allain {et~al.}(1996)Allain, Leach, \& Sedlmayr}]{Allain1996}
Allain, T., Leach, S., \& Sedlmayr, E. 1996, A\&A, 305, 616

\bibitem[{{\'A}lvarez-Barcia {et~al.}(2018){\'A}lvarez-Barcia, Russ,
  K{\"a}stner, \& Lamberts}]{Alvarez2018}
{\'A}lvarez-Barcia, S., Russ, P., K{\"a}stner, J., \& Lamberts, T. 2018,
  Monthly Notices of the Royal Astronomical Society, 479, 2007

\bibitem[{Anderson {et~al.}(2017)Anderson, Bergin, Blake, Ciesla, Visser, \&
  Lee}]{Anderson2017}
Anderson, D.~E., Bergin, E.~A., Blake, G.~A., {et~al.} 2017, The Astrophysical
  Journal, 845, 13

\bibitem[{Andres {et~al.}(1998)Andres, Domingo, Picher, \& Safont}]{Andres1998}
Andres, J., Domingo, L., Picher, M., \& Safont, V. 1998, International journal
  of quantum chemistry, 66, 9

\bibitem[{Apeloig(1990)}]{Apeloig1990}
Apeloig, Y. 1990, The Chemistry of Enols, ed. Z. Rappoport (Wiley,
  Chichester:John Wiley \& Sons)

\bibitem[{{Bacmann} {et~al.}(2012){Bacmann}, {Taquet}, {Faure}, {Kahane}, \&
  {Ceccarelli}}]{Bacmann2012}
{Bacmann}, A., {Taquet}, V., {Faure}, A., {Kahane}, C., \& {Ceccarelli}, C.
  2012, \aap, 541, L12, \dodoi{10.1051/0004-6361/201219207}

\bibitem[{{Bisschop} {et~al.}(2008){Bisschop}, {J{\o}rgensen}, {Bourke},
  {Bottinelli}, \& {van Dishoeck}}]{Bisschop2008}
{Bisschop}, S.~E., {J{\o}rgensen}, J.~K., {Bourke}, T.~L., {Bottinelli}, S., \&
  {van Dishoeck}, E.~F. 2008, \aap, 488, 959,
  \dodoi{10.1051/0004-6361:200809673}

\bibitem[{{Boogert} {et~al.}(2015){Boogert}, {Gerakines}, \&
  {Whittet}}]{Boogert2015}
{Boogert}, A.~C.~A., {Gerakines}, P.~A., \& {Whittet}, D.~C.~B. 2015, \araa,
  53, 541, \dodoi{10.1146/annurev-astro-082214-122348}

\bibitem[{Bosco {et~al.}(1984)Bosco, Nava, Brobst, \& Stief}]{Bosco1984}
Bosco, S., Nava, D., Brobst, W., \& Stief, L. 1984, The Journal of chemical
  physics, 81, 3505

\bibitem[{Boudin {et~al.}(1998)Boudin, Schutte, Greenberg,
  {et~al.}}]{Boudin1998}
Boudin, N., Schutte, W.~A., Greenberg, J.~M., {et~al.} 1998, A\&A, 331, 749

\bibitem[{Brooke {et~al.}(1996)Brooke, Tokunaga, Weaver, Crovisier,
  Bockel{\'e}e-Morvan, \& Crisp}]{Brooke1996}
Brooke, T., Tokunaga, A., Weaver, H., {et~al.} 1996, Nature, 383, 606

\bibitem[{Brunning \& Stief(1985)}]{Brunning1985}
Brunning, J., \& Stief, L. 1985, The Journal of chemical physics, 83, 1005

\bibitem[{{Butscher} {et~al.}(2015){Butscher}, {Duvernay}, {Theule}, {Danger},
  {Carissan}, {Hagebaum-Reignier}, \& {Chiavassa}}]{Butscher2015}
{Butscher}, T., {Duvernay}, F., {Theule}, P., {et~al.} 2015, \mnras, 453, 1587,
  \dodoi{10.1093/mnras/stv1706}

\bibitem[{{Canta} {et~al.}(2023){Canta}, {{\"O}berg}, \&
  {Rajappan}}]{Canta2023}
{Canta}, A., {{\"O}berg}, K.~I., \& {Rajappan}, M. 2023, \apj, 953, 81,
  \dodoi{10.3847/1538-4357/acda99}

\bibitem[{Carr \& Najita(2008)}]{Carr2008}
Carr, J.~S., \& Najita, J.~R. 2008, Science, 319, 1504

\bibitem[{Cecchi-Pestellini \& Aiello(1992)}]{Cecchi1992}
Cecchi-Pestellini, C., \& Aiello, S. 1992, Monthly Notices of the Royal
  Astronomical Society, 258, 125

\bibitem[{{Cernicharo} {et~al.}(2012){Cernicharo}, {Marcelino}, {Roueff},
  {Gerin}, {Jim{\'e}nez-Escobar}, \& {Mu{\~n}oz Caro}}]{Cernicharo2012}
{Cernicharo}, J., {Marcelino}, N., {Roueff}, E., {et~al.} 2012, ApJL, 759, L43,
  \dodoi{10.1088/2041-8205/759/2/L43}

\bibitem[{Cernicharo {et~al.}(1999)Cernicharo, Yamamura, Gonz{\'a}lez-Alfonso,
  de~Jong, Heras, Escribano, \& Ortigoso}]{Cernicharo1999}
Cernicharo, J., Yamamura, I., Gonz{\'a}lez-Alfonso, E., {et~al.} 1999, ApJL,
  526, L41

\bibitem[{Charnley(2004)}]{Charnley2004}
Charnley, S. 2004, Advances in Space Research, 33, 23

\bibitem[{{Charnley} {et~al.}(2001){Charnley}, {Rodgers}, \&
  {Ehrenfreund}}]{Charnley2001}
{Charnley}, S.~B., {Rodgers}, S.~D., \& {Ehrenfreund}, P. 2001, \aap, 378,
  1024, \dodoi{10.1051/0004-6361:20011193}

\bibitem[{Chevance {et~al.}(2020)Chevance, Kruijssen, Vazquez-Semadeni,
  Nakamura, Klessen, Ballesteros-Paredes, Inutsuka, Adamo, \&
  Hennebelle}]{Chevance2020}
Chevance, M., Kruijssen, J.~D., Vazquez-Semadeni, E., {et~al.} 2020, Space
  Science Reviews, 216

\bibitem[{Chuang {et~al.}(2020)Chuang, Fedoseev, Qasim, Ioppolo, J{\"a}ger,
  Henning, Palumbo, van Dishoeck, \& Linnartz}]{Chuang2020}
Chuang, K., Fedoseev, G., Qasim, D., {et~al.} 2020, Astronomy \& Astrophysics,
  635, A199

\bibitem[{Chuang(2018)}]{Chuang2018thesis}
Chuang, K.~J. 2018, PhD thesis, Leiden University

\bibitem[{{Chuang} {et~al.}(2016){Chuang}, {Fedoseev}, {Ioppolo}, {van
  Dishoeck}, \& {Linnartz}}]{Chuang2016}
{Chuang}, K.-J., {Fedoseev}, G., {Ioppolo}, S., {van Dishoeck}, E.~F., \&
  {Linnartz}, H. 2016, \mnras, 455, 1702, \dodoi{10.1093/mnras/stv2288}

\bibitem[{{Chuang} {et~al.}(2017){Chuang}, {Fedoseev}, {Qasim}, {Ioppolo}, {van
  Dishoeck}, \& {Linnartz}}]{Chuang2017}
{Chuang}, K.-J., {Fedoseev}, G., {Qasim}, D., {et~al.} 2017, \mnras, 467, 2552,
  \dodoi{10.1093/mnras/stx222}

\bibitem[{Chuang {et~al.}(2021)Chuang, Fedoseev, Scir{\`e}, Baratta, J{\"a}ger,
  Henning, Linnartz, \& Palumbo}]{Chuang2021}
Chuang, K.-J., Fedoseev, G., Scir{\`e}, C., {et~al.} 2021, Astronomy \&
  Astrophysics, 650, A85

\bibitem[{Chuang {et~al.}(2022)Chuang, J{\"a}ger, Krasnokutski, Fulvio, \&
  Henning}]{Chuang2022}
Chuang, K.-J., J{\"a}ger, C., Krasnokutski, S., Fulvio, D., \& Henning, T.
  2022, The Astrophysical Journal, 933, 107

\bibitem[{Collin \& Delwiche(1967)}]{Collin1967}
Collin, J.~E., \& Delwiche, J. 1967, Canadian Journal of Chemistry, 45, 1883

\bibitem[{Compagnini {et~al.}(2009)Compagnini, D’Urso, Puglisi, Baratta, \&
  Strazzulla}]{Compagnini2009}
Compagnini, G., D’Urso, L., Puglisi, O., Baratta, G., \& Strazzulla, G. 2009,
  Carbon, 47, 1605

\bibitem[{Dickerson {et~al.}(2018)Dickerson, Bera, \& Lee}]{Dickerson2018}
Dickerson, C.~E., Bera, P.~P., \& Lee, T.~J. 2018, The Journal of Physical
  Chemistry A, 122, 8898

\bibitem[{Drozdovskaya {et~al.}(2019)Drozdovskaya, van Dishoeck, Rubin,
  J{\o}rgensen, \& Altwegg}]{Drozdovskaya2019}
Drozdovskaya, M.~N., van Dishoeck, E.~F., Rubin, M., J{\o}rgensen, J.~K., \&
  Altwegg, K. 2019, Monthly Notices of the Royal Astronomical Society, 490, 50

\bibitem[{Evans {et~al.}(2012)Evans, Yu, Roberts, Stavros, \&
  Ullrich}]{Evans2012}
Evans, N.~L., Yu, H., Roberts, G.~M., Stavros, V.~G., \& Ullrich, S. 2012,
  Physical Chemistry Chemical Physics, 14, 10401

\bibitem[{{Fedoseev} {et~al.}(2017){Fedoseev}, {Chuang}, {Ioppolo}, {Qasim},
  {van Dishoeck}, \& {Linnartz}}]{Fedoseev2017}
{Fedoseev}, G., {Chuang}, K.-J., {Ioppolo}, S., {et~al.} 2017, ApJ, 842, 52,
  \dodoi{10.3847/1538-4357/aa74dc}

\bibitem[{{Fedoseev} {et~al.}(2015){Fedoseev}, {Ioppolo}, \&
  {Linnartz}}]{Fedoseev2015a}
{Fedoseev}, G., {Ioppolo}, S., \& {Linnartz}, H. 2015, \mnras, 446, 449,
  \dodoi{10.1093/mnras/stu1852}

\bibitem[{Ferris \& Ishikawa(1988)}]{Ferris1988}
Ferris, J.~P., \& Ishikawa, Y. 1988, Journal of the American Chemical Society,
  110, 4306

\bibitem[{Georgieva \& Velcheva(2006)}]{Georgieva2006}
Georgieva, M.~K., \& Velcheva, E.~A. 2006, International journal of quantum
  chemistry, 106, 1316

\bibitem[{Gerakines {et~al.}(2004)Gerakines, Moore, \& Hudson}]{Gerakines2004}
Gerakines, P., Moore, M., \& Hudson, R. 2004, Icarus, 170, 202

\bibitem[{Gerakines {et~al.}(2022)Gerakines, Yarnall, \&
  Hudson}]{Gerakines2022}
Gerakines, P.~A., Yarnall, Y.~Y., \& Hudson, R.~L. 2022, Monthly Notices of the
  Royal Astronomical Society, 509, 3515

\bibitem[{{Hama} \& {Watanabe}(2013)}]{Hama2013}
{Hama}, T., \& {Watanabe}, N. 2013, Chemical Reviews, 113, 8783,
  \dodoi{10.1021/cr4000978}

\bibitem[{Hamada {et~al.}(1984)Hamada, Hashiguchi, Tsuboi, Koga, \&
  Kondo}]{Hamada1984}
Hamada, Y., Hashiguchi, K., Tsuboi, M., Koga, Y., \& Kondo, S. 1984, Journal of
  Molecular Spectroscopy, 105, 93

\bibitem[{Hashiguchi {et~al.}(1984)Hashiguchi, Hamada, Tsuboi, Koga, \&
  Kondo}]{Hashiguchi1984}
Hashiguchi, K., Hamada, Y., Tsuboi, M., Koga, Y., \& Kondo, S. 1984, Journal of
  Molecular Spectroscopy, 105, 81

\bibitem[{Hatch {et~al.}(1992)Hatch, Polizzotti, Dougal, \&
  Rabinowitz}]{Hatch1992}
Hatch, S., Polizzotti, R., Dougal, S., \& Rabinowitz, P. 1992, Chemical physics
  letters, 196, 97

\bibitem[{Hattori {et~al.}(2005)Hattori, Suzuki, \& Shimizu}]{Hattori2005}
Hattori, R., Suzuki, E., \& Shimizu, K. 2005, Journal of molecular structure,
  750, 123

\bibitem[{He {et~al.}(2022)He, Simons, Fedoseev, Chuang, Qasim, Lamberts,
  Ioppolo, McGuire, Cuppen, \& Linnartz}]{He2022}
He, J., Simons, M., Fedoseev, G., {et~al.} 2022, Astronomy \& Astrophysics,
  659, A65

\bibitem[{Hennig \& Wagner(1995)}]{Hennig1995}
Hennig, G., \& Wagner, H. 1995, Berichte der Bunsengesellschaft f{\"u}r
  physikalische Chemie, 99, 989

\bibitem[{{Herbst} \& {van Dishoeck}(2009)}]{Herbst2009}
{Herbst}, E., \& {van Dishoeck}, E.~F. 2009, \araa, 47, 427,
  \dodoi{10.1146/annurev-astro-082708-101654}

\bibitem[{{Hiraoka} {et~al.}(2000){Hiraoka}, {Takayama}, {Euchi}, {Hand a}, \&
  {Sato}}]{Hiraoka2000}
{Hiraoka}, K., {Takayama}, T., {Euchi}, A., {Hand a}, H., \& {Sato}, T. 2000,
  ApJ, 532, 1029, \dodoi{10.1086/308612}

\bibitem[{Hudson {et~al.}(2014)Hudson, Ferrante, \& Moore}]{Hudson2014}
Hudson, R., Ferrante, R., \& Moore, M. 2014, Icarus, 228, 276

\bibitem[{Hudson \& Moore(1997)}]{Hudson1997}
Hudson, R., \& Moore, M. 1997, Icarus, 126, 233

\bibitem[{Hudson \& Moore(2004)}]{Hudson2004}
---. 2004, Icarus, 172, 466

\bibitem[{Hudson {et~al.}(2022)Hudson, Gerakines, \& Yarnall}]{Hudson2022}
Hudson, R.~L., Gerakines, P.~A., \& Yarnall, Y.~Y. 2022, The Astrophysical
  Journal, 925, 156

\bibitem[{Ito \& Nakanaga(2010)}]{Ito2010}
Ito, F., \& Nakanaga, T. 2010, Journal of Molecular Spectroscopy, 264, 100

\bibitem[{Jacox(1979)}]{Jacox1979}
Jacox, M.~E. 1979, Chemical Physics, 43, 157

\bibitem[{Jochims {et~al.}(1994)Jochims, Ruhl, Baumgartel, Tobita, \&
  Leach}]{Jochims1994}
Jochims, H., Ruhl, E., Baumgartel, H., Tobita, S., \& Leach, S. 1994, ApJ, 420,
  307

\bibitem[{{J{\o}rgensen} {et~al.}(2012){J{\o}rgensen}, {Favre}, {Bisschop},
  {Bourke}, {van Dishoeck}, \& {Schmalzl}}]{Jorgensen2012}
{J{\o}rgensen}, J.~K., {Favre}, C., {Bisschop}, S.~E., {et~al.} 2012, ApJL,
  757, L4, \dodoi{10.1088/2041-8205/757/1/L4}

\bibitem[{{J{\o}rgensen} {et~al.}(2016){J{\o}rgensen}, {van der Wiel},
  {Coutens}, {Lykke}, {M{\"u}ller}, {van Dishoeck}, {Calcutt}, {Bjerkeli},
  {Bourke}, {Drozdovskaya}, {Favre}, {Fayolle}, {Garrod}, {Jacobsen},
  {{\"O}berg}, {Persson}, \& {Wampfler}}]{Jorgensen2016}
{J{\o}rgensen}, J.~K., {van der Wiel}, M.~H.~D., {Coutens}, A., {et~al.} 2016,
  \aap, 595, A117, \dodoi{10.1051/0004-6361/201628648}

\bibitem[{Kameneva {et~al.}(2017)Kameneva, Volosatova, \&
  Feldman}]{Kameneva2017}
Kameneva, S.~V., Volosatova, A.~D., \& Feldman, V.~I. 2017, Radiation Physics
  and Chemistry, 141, 363

\bibitem[{Kamp {et~al.}(2023)Kamp, Henning, Arabhavi, Bettoni, Christiaens,
  Gasman, Grant, Morales-Calder{\'o}n, Tabone, Abergel, {et~al.}}]{Kamp2023}
Kamp, I., Henning, T., Arabhavi, A.~M., {et~al.} 2023, Faraday discussions,
  245, 112

\bibitem[{Kaye \& Strobel(1983)}]{Kaye1983}
Kaye, J.~A., \& Strobel, D.~F. 1983, Icarus, 54, 417

\bibitem[{Knez {et~al.}(2012)Knez, Moore, Ferrante, \& Hudson}]{Knez2012}
Knez, C., Moore, M., Ferrante, R., \& Hudson, R. 2012, ApJ, 748, 95

\bibitem[{Knez {et~al.}(2008)Knez, Moore, Travis, Ferrante, Chiar, Boogert,
  Mundy, Pendleton, Tielens, van Dishoeck, {et~al.}}]{Knez2008}
Knez, C., Moore, M., Travis, S., {et~al.} 2008, Proceedings of the
  International Astronomical Union, 4, 47

\bibitem[{Kobayashi {et~al.}(2017)Kobayashi, Hidaka, Lamberts, Hama, Kawakita,
  K{\"a}stner, \& Watanabe}]{Kobayashi2017}
Kobayashi, H., Hidaka, H., Lamberts, T., {et~al.} 2017, ApJ, 837, 155

\bibitem[{Kress {et~al.}(2010)Kress, Tielens, \& Frenklach}]{Kress2010}
Kress, M.~E., Tielens, A.~G., \& Frenklach, M. 2010, Advances in Space
  Research, 46, 44

\bibitem[{Lacy {et~al.}(1989)Lacy, Evans, Achtermann, Bruce, Arens, \&
  Carr}]{Lacy1989}
Lacy, J., Evans, N.~J., Achtermann, J., {et~al.} 1989, ApJ, 342, L43

\bibitem[{Lahuis \& van Dishoeck(2000)}]{Lahuis2000}
Lahuis, F., \& van Dishoeck, E. 2000, A\&A, 355, 699

\bibitem[{Lammertsma \& Prasad(1994)}]{Lammertsma1994}
Lammertsma, K., \& Prasad, B.~V. 1994, Journal of the American Chemical
  Society, 116, 642

\bibitem[{Lesclaux {et~al.}(1985)Lesclaux, Veyret, \& Roussel}]{Lesclaux1985}
Lesclaux, R., Veyret, B., \& Roussel, P. 1985, Berichte der Bunsengesellschaft
  f{\"u}r physikalische Chemie, 89, 330

\bibitem[{Lin {et~al.}(1995)Lin, Wu, \& Lien}]{Lin1995}
Lin, J.-F., Wu, C.-C., \& Lien, M.-H. 1995, The Journal of Physical Chemistry,
  99, 16903

\bibitem[{Linnartz {et~al.}(2015)Linnartz, Ioppolo, \& Fedoseev}]{Linnartz2015}
Linnartz, H., Ioppolo, S., \& Fedoseev, G. 2015, International Reviews in
  Physical Chemistry, 34, 205

\bibitem[{Lo {et~al.}(2020)Lo, Peng, Chou, Lu, \& Cheng}]{Lo2020}
Lo, J.-I., Peng, Y.-C., Chou, S.-L., Lu, H.-C., \& Cheng, B.-M. 2020, MNRAS,
  499, 543

\bibitem[{Locht {et~al.}(1991)Locht, Leyh, Denzer, Hagenow, \&
  Baumg{\"a}rtel}]{Locht1991}
Locht, R., Leyh, B., Denzer, W., Hagenow, G., \& Baumg{\"a}rtel, H. 1991,
  Chemical physics, 155, 407

\bibitem[{Loomis {et~al.}(2013)Loomis, Zaleski, Steber, Neill, Muckle, Harris,
  Hollis, Jewell, Lattanzi, Lovas, {et~al.}}]{Loomis2013}
Loomis, R.~A., Zaleski, D.~P., Steber, A.~L., {et~al.} 2013, The Astrophysical
  Journal Letters, 765, L9

\bibitem[{Lovas {et~al.}(2006)Lovas, Hollis, Remijan, \& Jewell}]{Lovas2006}
Lovas, F.~J., Hollis, J., Remijan, A.~J., \& Jewell, P. 2006, The Astrophysical
  Journal, 645, L137

\bibitem[{Mart{\'\i}n-Dom{\'e}nech {et~al.}(2020)Mart{\'\i}n-Dom{\'e}nech,
  {\"O}berg, \& Rajappan}]{Martin2020}
Mart{\'\i}n-Dom{\'e}nech, R., {\"O}berg, K.~I., \& Rajappan, M. 2020, ApJ, 894,
  98

\bibitem[{McGuire(2018)}]{McGuire2018}
McGuire, B.~A. 2018, The Astrophysical Journal Supplement Series, 239, 17

\bibitem[{McNaughton \& Evans(1999)}]{McNaughton1999}
McNaughton, D., \& Evans, C.~J. 1999, Journal of molecular spectroscopy, 196,
  274

\bibitem[{Mencos \& Krim(2016)}]{Mencos2016}
Mencos, A., \& Krim, L. 2016, Monthly Notices of the Royal Astronomical
  Society, 460, 1990

\bibitem[{Miller \& Klippenstein(2004)}]{Miller2004}
Miller, J.~A., \& Klippenstein, S.~J. 2004, Physical Chemistry Chemical
  Physics, 6, 1192

\bibitem[{Molpeceres \& Rivilla(2022)}]{Molpeceres2022}
Molpeceres, G., \& Rivilla, V.~M. 2022, Astronomy \& Astrophysics, 665, A27

\bibitem[{Moskaleva \& Lin(1998)}]{Moskaleva1998}
Moskaleva, L., \& Lin, M.-C. 1998, The Journal of Physical Chemistry A, 102,
  4687

\bibitem[{Mumma {et~al.}(2003)Mumma, DiSanti, Russo, Magee-Sauer, Gibb, \&
  Novak}]{Mumma2003}
Mumma, M., DiSanti, M., Russo, N.~D., {et~al.} 2003, Advances in Space
  Research, 31, 2563

\bibitem[{Mungan {et~al.}(1991)Mungan, Spitzer, Sethna, \&
  Sievers}]{Mungan1991}
Mungan, C., Spitzer, R., Sethna, J., \& Sievers, A. 1991, Physical Review B,
  43, 43

\bibitem[{Noble {et~al.}(2013)Noble, Theule, Borget, Danger, Chomat, Duvernay,
  Mispelaer, \& Chiavassa}]{Noble2013}
Noble, J.~A., Theule, P., Borget, F., {et~al.} 2013, Monthly Notices of the
  Royal Astronomical Society, 428, 3262

\bibitem[{{{\"O}berg} {et~al.}(2010){{\"O}berg}, {Bottinelli}, {J{\o}rgensen},
  \& {van Dishoeck}}]{Oberg2010}
{{\"O}berg}, K.~I., {Bottinelli}, S., {J{\o}rgensen}, J.~K., \& {van Dishoeck},
  E.~F. 2010, ApJ, 716, 825, \dodoi{10.1088/0004-637X/716/1/825}

\bibitem[{Okabe(1975)}]{Okabe1975}
Okabe, H. 1975, The Journal of Chemical Physics, 62, 2782

\bibitem[{Okabe \& Lenzi(1967)}]{Okabe1967}
Okabe, H., \& Lenzi, M. 1967, The Journal of Chemical Physics, 47, 5241

\bibitem[{Padovani {et~al.}(2024)Padovani, Galli, Scarlett, Grassi, Rehill,
  Zammit, Bray, \& Fursa}]{Padovani2024}
Padovani, M., Galli, D., Scarlett, L.~H., {et~al.} 2024, Astronomy \&
  Astrophysics, 682, A131

\bibitem[{Pereira {et~al.}(2020)Pereira, de~Barros, da~Costa, Oliveira, Fulvio,
  \& da~Silveira}]{Pereira2020}
Pereira, R., de~Barros, A., da~Costa, C., {et~al.} 2020, MNRAS, 495, 40

\bibitem[{Perrero {et~al.}(2022)Perrero, Enrique-Romero, Mart{\'\i}nez-Bachs,
  Ceccarelli, Balucani, Ugliengo, \& Rimola}]{Perrero2022}
Perrero, J., Enrique-Romero, J., Mart{\'\i}nez-Bachs, B., {et~al.} 2022, ACS
  Earth and Space Chemistry, 6, 496

\bibitem[{Potapov {et~al.}(2019)Potapov, J{\"a}ger, \& Henning}]{Potapov2019}
Potapov, A., J{\"a}ger, C., \& Henning, T. 2019, ApJ, 880, 12

\bibitem[{{Prasad} \& {Tarafdar}(1983)}]{Prasad1983}
{Prasad}, S.~S., \& {Tarafdar}, S.~P. 1983, ApJ, 267, 603,
  \dodoi{10.1086/160896}

\bibitem[{Raczy{\'n}ska {et~al.}(2005)Raczy{\'n}ska, Kosi{\'n}ska,
  O{\'s}mia{\l}owski, \& Gawinecki}]{Raczynska2005}
Raczy{\'n}ska, E.~D., Kosi{\'n}ska, W., O{\'s}mia{\l}owski, B., \& Gawinecki,
  R. 2005, Chemical reviews, 105, 3561

\bibitem[{{Shen} {et~al.}(2004){Shen}, {Greenberg}, {Schutte}, \& {van
  Dishoeck}}]{Shen2004}
{Shen}, C.~J., {Greenberg}, J.~M., {Schutte}, W.~A., \& {van Dishoeck}, E.~F.
  2004, \aap, 415, 203, \dodoi{10.1051/0004-6361:20031669}

\bibitem[{Simons {et~al.}(2020)Simons, Lamberts, \& Cuppen}]{Simons2020}
Simons, M., Lamberts, T., \& Cuppen, H. 2020, A\&A, 634, A52

\bibitem[{Stolkin {et~al.}(1977)Stolkin, Ha, \& G{\"u}nthard}]{Stolkin1977}
Stolkin, I., Ha, T.-K., \& G{\"u}nthard, H.~H. 1977, Chemical physics, 21, 327

\bibitem[{Tabone {et~al.}(2023)Tabone, Bettoni, van Dishoeck, Arabhavi, Grant,
  Gasman, Henning, Kamp, G{\"u}del, Lagage, {et~al.}}]{Tabone2023}
Tabone, B., Bettoni, G., van Dishoeck, E., {et~al.} 2023, Nature Astronomy, 1

\bibitem[{Taquet {et~al.}(2017)Taquet, Wirstr{\"o}m, Charnley, Faure,
  L{\'o}pez-Sepulcre, \& Persson}]{Taquet2017}
Taquet, V., Wirstr{\"o}m, E., Charnley, S.~B., {et~al.} 2017, A\&A, 607, A20

\bibitem[{Thiel {et~al.}(2017)Thiel, Belloche, Menten, Garrod, \&
  M{\"u}ller}]{Thiel2017}
Thiel, V., Belloche, A., Menten, K., Garrod, R., \& M{\"u}ller, H. 2017,
  Astronomy \& Astrophysics, 605, L6

\bibitem[{Tielens(2013)}]{Tielens2013}
Tielens, A. 2013, Reviews of Modern Physics, 85, 1021

\bibitem[{{Tielens}(1992)}]{Tielens1992}
{Tielens}, A. G. G.~M. 1992, in Tokyo: Univ. Tokyo Press, Vol. 251, Chemistry
  and Spectroscopy of Interstellar Molecules, ed. N.~{Kaifu}, 237

\bibitem[{Tsuge \& Watanabe(2023)}]{Tsuge2023}
Tsuge, M., \& Watanabe, N. 2023, Proceedings of the Japan Academy, Series B,
  99, 103

\bibitem[{Turner {et~al.}(2021)Turner, Chandra, Fortenberry, \&
  Kaiser}]{Turner2021}
Turner, A.~M., Chandra, S., Fortenberry, R.~C., \& Kaiser, R.~I. 2021,
  ChemPhysChem, 22, 985

\bibitem[{van Dishoeck {et~al.}(2023)van Dishoeck, Grant, Tabone, van Gelder,
  Francis, Tychoniec, Bettoni, Arabhavi, Gasman, Kavanagh,
  {et~al.}}]{vanDishoeck2023}
van Dishoeck, E., Grant, S., Tabone, B., {et~al.} 2023, Faraday Discussions

\bibitem[{van Gelder {et~al.}(2020)van Gelder, Tabone, van Dishoeck, Beuther,
  Boogert, o~Garatti, Klaassen, Linnartz, M{\"u}ller, Taquet,
  {et~al.}}]{vanGelder2020}
van Gelder, M., Tabone, B., van Dishoeck, E., {et~al.} 2020, A\&A, 639, A87

\bibitem[{Vazart {et~al.}(2020)Vazart, Ceccarelli, Balucani, Bianchi, \&
  Skouteris}]{Vazart2020}
Vazart, F., Ceccarelli, C., Balucani, N., Bianchi, E., \& Skouteris, D. 2020,
  Monthly Notices of the Royal Astronomical Society, 499, 5547

\bibitem[{Vinogradoff {et~al.}(2012)Vinogradoff, Duvernay, Farabet, Danger,
  Theul{\'e}, Borget, Guillemin, \& Chiavassa}]{Vinogradoff2012}
Vinogradoff, V., Duvernay, F., Farabet, M., {et~al.} 2012, The Journal of
  Physical Chemistry A, 116, 2225

\bibitem[{Volosatova {et~al.}(2022)Volosatova, Zasimov, \&
  Feldman}]{Volosatova2022}
Volosatova, A.~D., Zasimov, P.~V., \& Feldman, V.~I. 2022, The Journal of
  Chemical Physics, 157

\bibitem[{West {et~al.}(2018)West, Castillo, Sit, Mohamad, Lowe, Joblin, Bodi,
  \& Mayer}]{West2018}
West, B., Castillo, S.~R., Sit, A., {et~al.} 2018, Physical Chemistry Chemical
  Physics, 20, 7195

\bibitem[{Wu {et~al.}(2002)Wu, Judge, Cheng, Shih, Yih, \& Ip}]{Wu2002}
Wu, C.~R., Judge, D., Cheng, B.-M., {et~al.} 2002, Icarus, 156, 456

\bibitem[{Xia {et~al.}(1991)Xia, Chien, Wu, \& Judge}]{Xia1991}
Xia, T., Chien, T., Wu, C.~R., \& Judge, D. 1991, Journal of Quantitative
  Spectroscopy and Radiative Transfer, 45, 77

\bibitem[{Zeng {et~al.}(2021)Zeng, Jim{\'e}nez-Serra, Rivilla,
  Mart{\'\i}n-Pintado, Rodr{\'\i}guez-Almeida, Tercero, de~Vicente,
  Rico-Villas, Colzi, Mart{\'\i}n, {et~al.}}]{Zeng2021}
Zeng, S., Jim{\'e}nez-Serra, I., Rivilla, V.~M., {et~al.} 2021, The
  Astrophysical Journal Letters, 920, L27

\bibitem[{Zhang {et~al.}(2023)Zhang, Wang, Turner, Marks, Chandra, Fortenberry,
  \& Kaiser}]{Zhang2023}
Zhang, C., Wang, J., Turner, A.~M., {et~al.} 2023, The Astrophysical Journal,
  952, 132

\bibitem[{Zhen {et~al.}(2014)Zhen, Castellanos, Paardekooper, Linnartz, \&
  Tielens}]{Zhen2014}
Zhen, J., Castellanos, P., Paardekooper, D.~M., Linnartz, H., \& Tielens, A.~G.
  2014, ApJL, 797, L30

\end{thebibliography}
\bibliographystyle{aasjournal}

\appendix


\section{Infrared spectra of C$_2$H$_2$:NH$_3$ ice mixtures}\label{appendix_A}

\begin{figure}[h]
        \begin{center}
                \includegraphics[width=90mm]{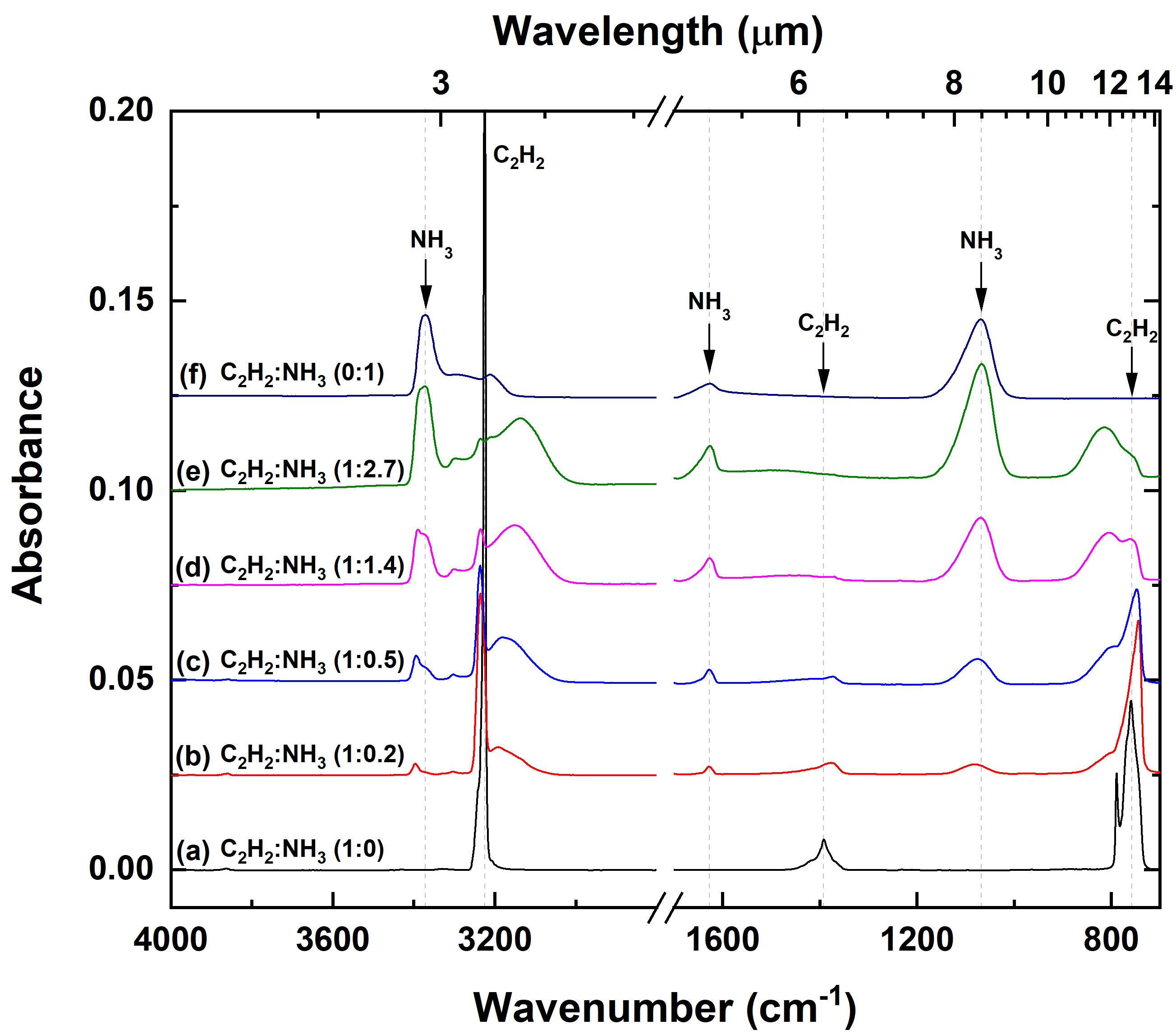}
                \caption{Infrared spectra of the deposited C$_2$H$_2$:NH$_3$ ice mixture with ratios of (a) 1:0, (b) 1:0.2, (c) 1:0.5, (d) 1:1.4, (e) 1:2.7, and (f) 0:1 at 10 K. The spectra are offset for clarity.}
                \label{FigA1}
        \end{center}
\end{figure}

Figure \ref{FigA1} shows the IR spectra obtained after the deposition of C$_2$H$_2$:NH$_3$ ice mixtures with ratios of (a) 1:0, (b) 1:0.2, (c) 1:0.5, (d) 1:1.4, (e) 1:2.7, and (f) 0:1 at 10 K. For a direct comparison, the absolute C$_2$H$_2$ abundance was kept the same, namely the averaged \textit{N}(C$_2$H$_2$)=(8$\pm$2)$\times$10$^{16}$ molecule cm$^{-2}$, in different ratios, while the column density \textit{N}(NH$_3$) changes from $\sim$2.3$\times$10$^{16}$ to $\sim$3.1$\times$10$^{17}$ molecule cm$^{2}$. The main features of pure C$_2$H$_2$ ice are found at 759/789, 1391, 1962, 3225 cm$^{-1}$ in the spectrum (a). It is clear that these IR peaks significantly shift and even broaden when C$_2$H$_2$ ice is mixed with NH$_3$ (i.e., spectra (b)-(e)). 
The non-negligible changes of C$_2$H$_2$ spectral characteristics have been reported in \cite{Knez2012} for CO, CO$_2$, CH$_4$, and H$_2$O ice. Future spectroscopic studies that include NH$_3$ ice are very much required and astronomically relevant to identifying C$_2$H$_2$ in the interstellar ice. 

\section{Kinetic evolution of IR absorbance area for the selected products}\label{appendix_B}

\begin{figure}[]
        \begin{center}
                \includegraphics[width=90mm]{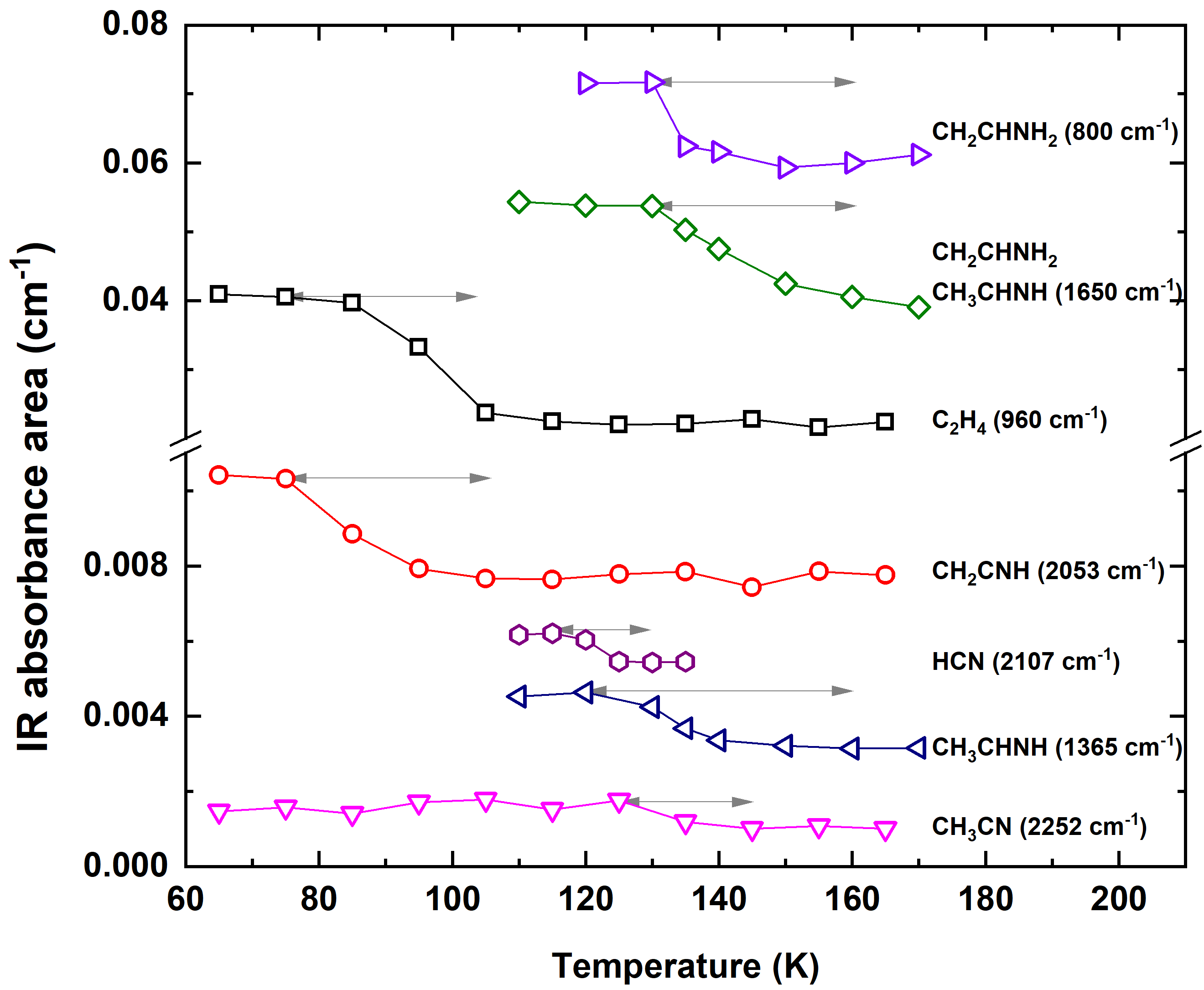}
                \caption{Kinetic evolution of the selected products as a function of temperature during the TPD experiment. The IR peaks used for monitoring the corresponding species are shown in parentheses. The arrows highlight the temperature range of desorption.}
                \label{FigA2}
        \end{center}
\end{figure}

Figure \ref{FigA2} presents the kinetic evolution of IR peak intensities for the selected products, including CH$_2$CHNH$_2$, CH$_3$CHNH, CH$_2$CNH, CH$_3$CN, C$_2$H$_4$, and HCN as a function of temperature during the TPD experiment with a ramping rate of 5 K min$^{-1}$. Each data point was collected over 5 K due to the IR measuring time (i.e., 60 seconds); the uncertainty of the recorded temperatures is $\sim$5 K.

\end{document}